\documentclass[11pt]{article}
\usepackage{comment,soul}
\usepackage[hidelinks]{hyperref}
\usepackage{enumitem}
\usepackage{amsmath,amssymb,epsf,cite,graphicx,subfigure}
\usepackage{amsmath,braket,tikzsymbols}
\usepackage[bbgreekl]{mathbbol}
\usepackage{comment}
\usepackage[export]{adjustbox}
\usepackage{dsfont}
\usepackage{faktor}
\usepackage{mathrsfs}
\usepackage{float}
\usepackage{empheq}

\numberwithin{equation}{section}

\definecolor{airforceblue}{rgb}{0.36, 0.54, 0.66}
\newcommand{\beq}{\begin{equation}}
\newcommand{\eeq}{\end{equation}}

\begin{document}
\baselineskip=15.5pt
\pagestyle{plain}
\setcounter{page}{1}

\begin{center}
{\LARGE \bf Norm of the no-boundary state}
\vskip 1cm

\textbf{Jordan Cotler$^{1,a}$ and Kristan Jensen$^{2,b}$}

\vspace{0.5cm}

{\it ${}^1$ Department of Physics, Harvard University, Cambridge, MA 02138, USA \\}
{\it ${}^2$ Department of Physics and Astronomy, University of Victoria, Victoria, BC V8W 3P6, Canada\\}

\vspace{0.3cm}

{\tt  ${}^a$jcotler@fas.harvard.edu, ${}^b$kristanj@uvic.ca\\}

\medskip

\end{center}

\vskip1cm

\begin{center}
{\bf Abstract}
\end{center}
\hspace{.3cm}
We consider Einstein gravity with positive cosmological constant coupled to matter in an asymptotically de Sitter universe with sphere boundary at timelike infinity. In this setting we show that, to one-loop order and at late time, the norm of the no-boundary state vanishes, going as $e^{S_0} \frac{Z S_0^{-d(d+1)/4}}{\text{vol}(SO(d,1))}$ with $S_0$ the tree-level entropy of the static patch, $d$ the spacetime dimension, and $Z$ non-negative.  We show that the presence of an observer stabilizes the norm to a large, positive value.

\newpage

\tableofcontents

\section{Introduction}

In this work we consider basic aspects of de Sitter quantum gravity at one-loop order, focusing on the no-boundary state, quantum mechanics at late time, and the Bekenstein-Hawking entropy associated with the horizon of the static patch. Throughout we work within effective field theory using the gravitational path integral. 

The no-boundary state was proposed long ago by Hartle and Hawking as a natural state of the universe in de Sitter space~\cite{Hartle:1983ai}. It is prepared by a Euclidean path integral in which spacetime caps off smoothly in imaginary time. The no-boundary state is theoretically natural since, in Einstein gravity with positive cosmological constant, it prepares a state invariant under the de Sitter isometries at the level of perturbation theory and so plays the role of a de Sitter ``vacuum.'' Moreover, the hypothesis that our universe is in the no-boundary state (and coupled to an inflaton) accurately predicts the observed spectrum of cosmic microwave background fluctuations~\cite{Baumann:2022mni, Maldacena:2024uhs}. However, this hypothesis also seems to predict far too small a universe to be consistent with a large inflationary epoch. (See~\cite{Maldacena:2024uhs} for a modern review of these successes and puzzles.)

In this paper we study formal properties of the no-boundary state in de Sitter quantum gravity.  We would like to better understand how the gravitational path integral produces genuine quantum states at late Lorentzian time, when space is large and gravitational perturbations are weak. We focus on the overlaps of states and expectation values therein, including one-loop corrections. We find surprising results on account of the de Sitter isometries and a proper treatment of the diffeomorphism invariance of gravity.

At tree level the most probable history of the universe in the no-boundary state is one in which a universe with a spherical spatial slice bubbles forth from Euclidean time and evolves to become global de Sitter space at late Lorentzian time. There are other subleading contributions to the no-boundary state, for example where there is a de Sitter universe at late time with torus spatial slice, but in this work we focus on the leading contribution.

For quantum field theory in this fixed spacetime, the analogue of the no-boundary state prepared by a Euclidean cap is the Bunch-Davies state. In that context the norm at late time is, by unitarity, equal to the norm at zero Lorentzian time, where it is given by the sphere partition function. That in turn is also equal to the thermal partition function for the field theory in the static patch of de Sitter at the Hawking temperature of the cosmological horizon. We review these results briefly in the next Section.

\begin{figure}[t!]
\begin{center}
\includegraphics[scale=.45]{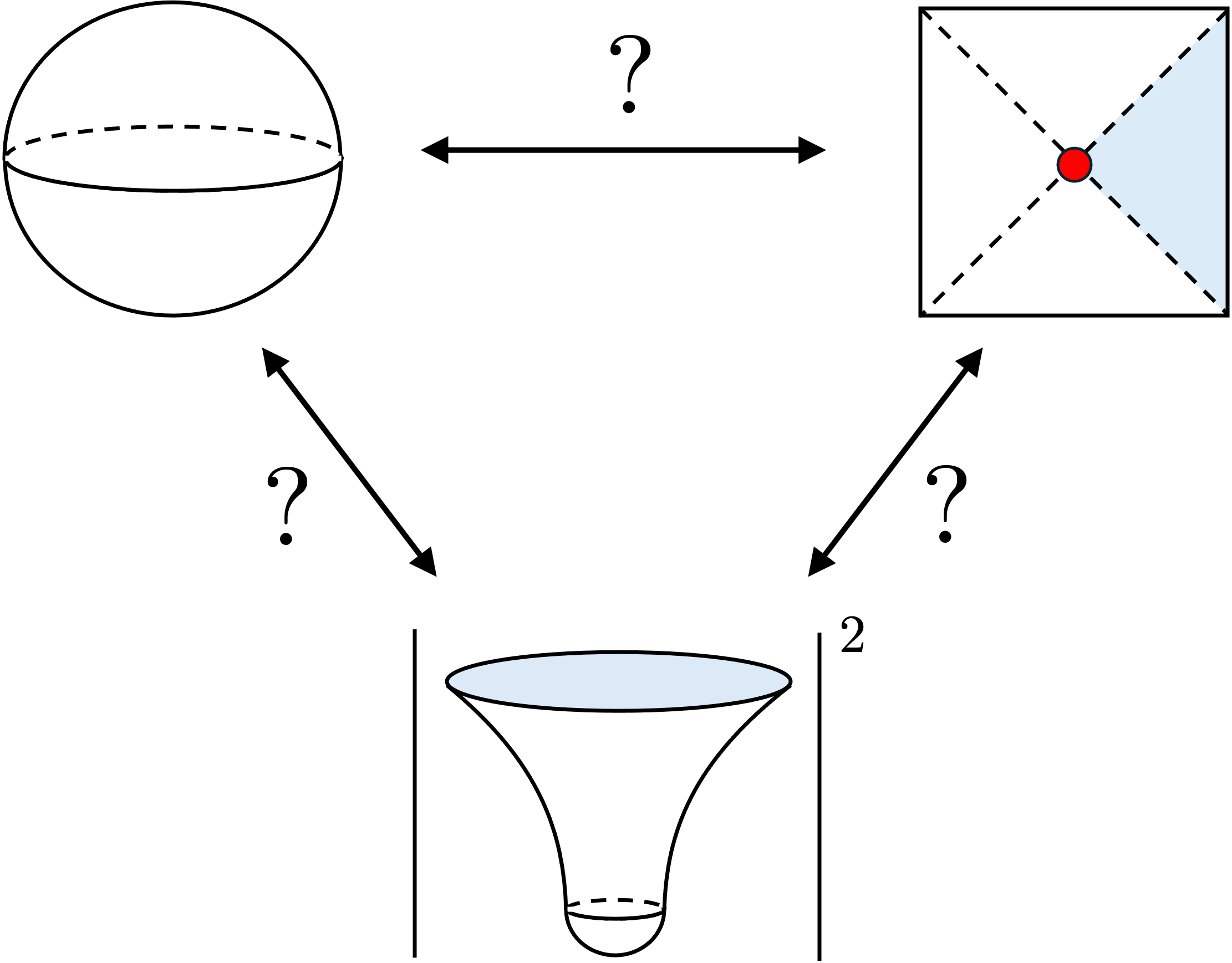}
\end{center}
\caption{Gibbons and Hawking proposed that the logarithm of the sphere partition function of a theory of de Sitter gravity is the horizon entropy of the static patch (indicated by the Penrose diagram of de Sitter in the upper right). We might also think that the sphere amplitude coincides with the norm of the no-boundary state at late time (whose wavefunction is computed by a path integral over geometries with a Euclidean cap represented at the bottom). \label{fig:conceptual}}
\end{figure}

Is there a similar story in gravity? By perturbative unitarity we might think that the late-time norm (or, at least, the contribution from a single universe with a spatial sphere) is equal to the sphere partition function of gravity, as for field theory. Gibbons and Hawking~\cite{Gibbons:1976ue} conjectured that in a theory of gravity the sphere partition function likewise gives a thermal partition function, but in the absence of a well-defined energy they proposed that it simply counts an entropy. A modern interpretation of their proposal is that the sphere partition function counts the dimension of the Hilbert space of states in the static patch (or perhaps $i$ times the dimension~\cite{Maldacena:2024spf}), and that this dimension is the exponential of the cosmological horizon entropy. See Fig.~\ref{fig:conceptual}. That proposal holds at tree level, however it cannot be correct quantum mechanically. The issue is that the sphere partition function of gravity has a phase at one loop except in $d \equiv 2\,\,(\text{mod}\,\,4)$ dimensions, as found by Polchinski~\cite{Polchinski:1988ua}. For example, the sphere partition function is imaginary in odd dimensions and negative in four, and thus betrays a state counting interpretation. 

Polchinski's phase has long been mysterious. A popular, conjectured interpretation is that the phase, analogous to that of multi-instanton amplitudes, signals an instability driving decay to a more probable spacetime. However, this claim has always been murky. In four dimensions, there is no known compact solution to the Einstein's equations that dominates over the sphere, and as emphasized by~\cite{Banihashemi:2024weu}, the mostly wrong-sign fluctuations of the conformal mode have yet to be connected to a property of the Lorentzian theory. (Although recently a similar phase has been related to real-time physics in the static patch of spacetimes that are the product of de Sitter space and a sphere~\cite{Turiaci:2025xwi, Shi:2025amq, Ivo:2025yek}.)

We are left with several puzzles. What is the physical meaning, if any, of Polchinski's phase in the real-time theory? In quantum field theory the sphere partition function is the norm of the Bunch-Davies state; in gravity, does Polchinski's phase indicate that the norm of the no-boundary state has a phase, i.e.~that it is non-normalizable? Can we even cut the sphere path integral in a theory of gravity and thereby interpret it as an overlap of states? If the loop-corrected static patch entropy is not given by the sphere partition function, is there a sensible definition at all, and if so, what is it? 

In this manuscript we propose answers to these questions using results we establish at late Lorentzian time. Let us first describe the results, and then our proposals. Our main result, valid to one-loop order in Newton's constant, is the late-time norm of a state in theories of $d$-dimensional Einstein gravity with positive cosmological constant, possibly coupled to matter. In this context ``late time'' means that the universe becomes asymptotically de Sitter, we have in mind wavefunctionals of the universe $\Psi[\gamma,\varphi]$ obtained from the gravitational path integral with suitable early time boundary conditions (like the no-boundary state), and $(\gamma,\varphi)$ refer to certain rescaled versions of the spatial metric and matter fields on a constant-time slice which remain finite in the $t\to\infty$ limit. Our results hold for states $\Psi$ which are peaked around $\gamma$ being the round metric on the sphere. Then the norm is 
\beq
\label{E:main}
	\langle \! \langle \Psi|\Psi \rangle \! \rangle = \int \frac{[d\gamma][d\varphi]}{\text{diff}\times\text{Weyl}} |\Psi[\gamma,\varphi]|^2\,,
\eeq
where the doubled ket $|\,\cdot\,\rangle \! \rangle$ denotes a state in quantum gravity.  Crucially, the wavefunction and Hilbert space of states at late time only depend on $(\gamma,\varphi)$ up to diff$\times$Weyl on the $t\to\infty$ slice, with the Weyl redundancy a vestige of bulk temporal diffeomorphism invariance. That feature mirrors the familiar fact in anti-de Sitter space that (up to anomalies) the bulk theory depends on the boundary data modulo boundary diffeomorphisms and Weyl rescalings.

The late-time overlap~\eqref{E:main} is natural in de Sitter gravity and has been proposed in~\cite{Chakraborty:2023los} and used in 3d gravity in \cite{Godet:2024ich,Collier:2025lux}. See in particular~\cite{Chakraborty:2023los} which studied correlation functions computed from the overlap. In this work we obtain~\eqref{E:main} from the gravitational path integral up to the constant $A$ that we are working to compute~\cite{CHJWIP2}, and exploit some features of the semiclassical no-boundary wavefunction to calculate its norm. While the authors of~\cite{Chakraborty:2023los} did not compute the norm of the no-boundary state using the overlap~\eqref{E:main}, our results are consistent with and complement theirs. 

Consider pure gravity in even $d>2$ for the moment. In $d=4$ the no-boundary wavefunction $\Psi_{\rm HH}[\gamma]$ has the feature that it is peaked around the round metric on the sphere, and we argue that this is true for $d>4$.\footnote{In $d=3$ the diff$\times$Weyl symmetry can fix $\gamma$ to the round metric so that the wavefunction becomes a constant.} Indeed, $|\Psi_{\rm HH}[\gamma]|^2$ takes the form of a right-sign Gaussian distribution in transverse traceless fluctuations away from the round sphere. The general result~\eqref{E:main} then implies that the late-time norm of the no-boundary state is, unlike the sphere amplitude, non-negative. However, in recasting the integral over metric fluctuations up to diff$\times$Weyl as one over transverse traceless fluctuations, we leave unfixed a residual gauge symmetry equal to the conformal isometry $SO(d,1)$ of the sphere. The one-loop contribution to the norm then involves a ratio of determinants over nonzero modes giving a finite prefactor after renormalization, divided by the volume of the residual gauge symmetry group $SO(d,1)$. As this group is noncompact we find the surprising result that the no-boundary state is null to one-loop order.  

In odd $d$ the wavefunction is not invariant under Weyl rescalings but transforms like an anomalous CFT partition function. For example, in $d=3$ the wavefunction behaves like the partition function of a CFT with an effective central charge $c = \frac{3L}{2G}i+13$ where $L$ is the de Sitter radius~\cite{Cotler:2019nbi}. In that case the integrand $|\Psi|^2$ behaves like a CFT partition function with $c_{\rm eff} = c + c^* = 26$, the magic number required for us to gauge diff$\times$Weyl symmetry without introducing any other field variables, i.e.~for which the expression~\eqref{E:main} is consistent where we only integrate over the spatial metric $\gamma$~\cite{Cotler:2019nbi, Collier:2025lux}. Effectively, the norm of the wavefunction and the measure each have Weyl anomalies that cancel, and we present an argument that this persists in general $d$. Furthermore, adding matter does not change these results in any dimension.

The $d=3$ case is particularly instructive, as then~\eqref{E:main} teaches us that there is an emergent bosonic worldsheet string theory at future infinity, with the computation of the norm presented here resembling that of the sphere amplitude of string theory with compact target. Indeed, the latter is an excellent mental model for our results, and many features suggested by it have a de Sitter analogue. For example, we expect that in computing late-time overlaps we should sum over topologies at future infinity. Indeed other contributions to the wavefunction, like when future infinity is a torus or a product of circles and spheres, generate nonzero contributions to the norm using~\eqref{E:main}. As a result the no-boundary state is not null, but rather its norm is only approximately zero to leading order in both the loop and topological expansions.

Besides adding handles, the analogy with the sphere amplitude in string theory suggests that we can obtain nonzero results by inserting ``vertex operators.'' That is, by studying unnormalized correlation functions of gauge-invariant observables through the overlap~\eqref{E:main}, we can effectively fix the conformal isometry. The authors of~\cite{Chakraborty:2023los} have proposed that correlation functions of this sort in flat slicing are the proper versions of cosmological correlators of gauge-fixed operators studied by theoretical cosmologists~\cite{Maldacena:2002vr,Weinberg:2005vy,Arkani-Hamed:2018kmz,Baumann:2022jpr}, with the details of the integration of data mod diff$\times$Weyl in the overlap~\eqref{E:main} being important for the normalization of low-point functions, as well as the gauge-invariance of four- and higher point functions.  Indeed, recall that in worldsheet string theory the two-point function of on-shell vertex operators is delicately finite, with a zero mode volume canceling against the volume of a residual $SO(1,1)$ boost symmetry that preserves the insertions~\cite{Erbin:2019uiz}, while three- and higher-point functions completely fix the noncompact part of the residual conformal invariance on the sphere. For de Sitter (in general dimensions), assuming suitable vertex operators exist, inserting two or more into the overlap~\eqref{E:main} leads to an unnormalized equal-time cosmological correlator, soaking up the noncompact residual symmetry by a similar argument. The precise gauge-invariant version of these operators is unknown for Einstein gravity coupled to matter in general dimension, and finding them is an important question to address in future work.

Another route to obtaining a finite norm is to let the ``vertex operators'' be the endpoints of worldlines joined through the Euclidean cap of the no-boundary state. We assume that the endpoints can be suitably dressed to give a gauge-invariant operator. We then study this case using a simple observer model for the worldline due to~\cite{Witten:2023xze}, defined with a continuous energy spectrum so that the observer entropy is infinite, and accrue evidence that this norm is large and positive, going as the exponential of the horizon entropy in the presence of the observer described by this worldline. A peculiar feature of our result is that the infinite observer entropy is important to cancel out the volume of a residual $SO(1,1)$ symmetry. An observer with finite entropy, such as that studied in~\cite{Maldacena:2024spf}, would not lead to a nonzero norm. The physics of this result is unclear. Perhaps observers which survive to the infinite future have `infinite resolution' and hence infinite entropy.

In the settings above, inserting field operators or endpoints of worldlines, the common feature is that we fix the residual conformal symmetry by studying a suitable unnormalized cosmological correlator. These insertions are required in order to achieve nonzero correlations at one-loop order.

With its one-loop approximation vanishing, what is the norm of the no-boundary state? As we discuss in Section~\ref{S:latetime}, the origin of the vanishing is due to two facts: (i) the one-loop inner product on asymptotic states has a residual $SO(d,1)$ gauge symmetry corresponding to conformal Killing vectors, and (ii) the one-loop Hartle-Hawking wavefunction (and also the Bunch-Davies state if there is matter involved) is invariant under $SO(d,1)$ and yet has no bosonic zero modes.  Then computing the inner product of (ii) using (i), upon fixing the residual gauge symmetry we find that the norm goes as one over a divergent volume.  For $d = 3$, we can argue directly that (i) and (ii) persist to all orders in perturbation theory, and so the sphere contribution to the norm is zero.  For $d > 3$, we show that the inner product on asymptotic states has a residual infinite gauge symmetry at all loops in perturbation theory, and so fact (i) persists.  In Section~\ref{S:matter} we suggest that loop corrections to the Hartle-Hawking and Bunch-Davies states may serve to renormalize couplings in the wavefunction, and that the corrected wavefunction is symmetric under the higher-loop residual gauge symmetry; if this holds, then (ii) persists beyond the one-loop approximation. Taken together, this would suggest that the sphere contribution to the norm vanishes beyond one loop even in $d>3$. We emphasize that while we establish (i) to all loops, there is more work to be done to establish, or refute, (ii) beyond one loop.

In sum, we find that for theories of gravity the sphere partition function does not equal the late-time norm of the no-boundary state. The former is nonzero with a phase while the latter is non-negative but vanishes to at least one loop. Introducing an observer is not enough to remove Polchinski's phase from the sphere partition function~\cite{Maldacena:2024spf}, while the late-time norm remains non-negative but is now stabilized to a finite value at one loop. We summarize the schematic form of these results in Table~\ref{T:summary}.

\begin{table}
\begin{center}
\begin{tabular}{c|c|c}
 & $Z_{\rm sphere}$ & $\langle \!\langle\text{HH}|\text{HH}\rangle\!\rangle$ \\
 \hline
 without observer & $ i^{d+2} e^{S_0}\frac{Z S_0^{-\mathcal{D}_d/2} }{\text{vol}(SO(d+1))}$ & $ e^{S_0}\frac{\widetilde{Z}S_0^{-\mathcal{D}_d/2} }{\text{vol}(SO(d,1))}$ 
 \\
 \hline
 with observer & $- i\,e^{S_0}\frac{Z' S_0^{-(\mathcal{D}_{d-2}+1)/2} }{\text{vol}(SO(d-1))}e^{S_{\rm obs} -\beta_{\rm dS}m}$ & $ e^{S_0}\frac{\widetilde{Z}' S_0^{-(\mathcal{D}_{d-2}+1)/2} }{\text{vol}(SO(d-1))}e^{-\beta_{\rm dS}m}$
\end{tabular}
 \caption{\label{T:summary}  A summary of the one-loop renormalized sphere partition function and norm of the no-boundary state without and with an observer of mass $m\propto 1/G$ described by the worldline model of~\cite{Witten:2023xze}. Here $S_0$ is the tree-level static patch entropy, $S_{\rm obs}$ is the observer entropy (implicitly taken to infinity in our expression for the norm), $\mathcal{D}_d = \text{dim}(SO(d,1)) = \frac{d(d+1)}{2}$, and the constants $Z, Z'$, etc., are non-negative ratios of renormalized one-loop determinants. The volumes and powers of $S_0$ arise from residual gauge symmetries.}
\end{center}
\end{table}

The vanishing norm is a surprising result. To come to terms with it, or at least with the discrepancy with the phases, we also present a more physical argument that the norm is non-negative. The basic observation is as follows. Polchinski's phase arises from the fact that the rotated conformal mode behaves like a tachyonic scalar on the sphere, so that the sphere partition function is finite only upon rotating the fluctuations of low angular momentum; if the conformal mode was a bona fide scalar field, then this phase would be physical, indicating that the Bunch-Davies wavefunction is non-normalizable, a wrong-sign Gaussian in certain directions. However, the late-time wavefunction of gravity does not actually depend on the fluctuations of the conformal mode, but only on the transverse traceless fluctuations of the metric $\gamma$ at future infinity. As such the conformal mode contributes to the normalization of the late-time wavefunction and to overlaps, but is not ``exchanged'' in the overlap~\eqref{E:main}.

With all of this being said we are in a position to state our proposed answers to the vexing questions mentioned above. Our basic observation is that the sphere partition function and late-time norm are genuinely different quantities. The norm is positive-definite and on those grounds we hypothesize that it defines the static patch entropy. To be stabilized to a finite value at one loop order we require insertions that preserve the time translation symmetry of the static patch. This can be done when the insertion is two copies of an observer (with infinite entropy), located in complementary static patches, in which case this setup and the result qualitatively resembles that of~\cite{Chandrasekaran:2022cip} in their construction of an algebra of observables in the static patch.

At least for the no-boundary state with sphere boundary at future infinity, we do not find a signature of Polchinski's phase at late time. The no-boundary wavefunction is a complex but right-sign joint Gaussian distribution in fluctuations of the metric at future infinity, with no non-normalizable directions. Our results suggest that, rather than being the overlap of the no-boundary ket with the no-boundary bra, the sphere partition function is a pairing of the ket with itself. See also~\cite{Witten:2025ayw} for a similar, recent interpretation of the Euclidean gravitational path integral. The complex conjugate of the amplitude would then be interpreted as the pairing of the no-boundary bra with itself. However, we do not yet know how to perform a computation that would test this proposal, nor another version of the Euclidean gravity sphere partition function that agrees with the inner product of the bra and ket of the no-boundary state.

The remainder of this manuscript is organized as follows. In Section~\ref{S:background} we briefly review the basics of the sphere partition function in field theory and gravity. Then in Section~\ref{S:latetime} we present the argument for the overlap~\eqref{E:main} from de Sitter gravity, and elaborate on the ensuing norm of the no-boundary state. We discuss the stabilization of the norm by anchors in Section~\ref{S:matter}, and conclude with a discussion of several observations and open questions in Section~\ref{S:discussion}.

\section{Background}
\label{S:background}

In this Section we briefly review the sphere partition function $Z_{\rm sphere}$ of Einstein gravity with positive cosmological constant, possibly coupled to free matter. We first recall the argument of Gibbons and Hawking~\cite{Gibbons:1976ue} that identifies $\log Z_{\rm sphere}$ with the static patch entropy, and then the derivation of Polchinski's phase~\cite{Polchinski:1988ua} that undermines it.

\subsection{Static patch entropy}

Consider unit-radius global de Sitter space in $d$ dimensions described by
\beq
ds^2 = -dt^2 + \cosh^2(t) d\Omega_{d-1}^2\,, \qquad d\Omega_{d-1}^2 = d\theta^2 + \sin^2(\theta)d\Omega_{d-2}^2\,.
\eeq
Global de Sitter space lacks a globally defined timelike Killing vector. However, the causal development of any timelike geodesic is a ``static patch'' admitting a timelike Killing vector. There is no unique static patch, but rather a $\faktor{SO(d,1)}{SO(1,1)\times SO(d-1)}$ coset of static patches corresponding to the set of timelike geodesics. Using the de Sitter isometries we can fix the geodesic to sit at the north pole $\theta=0$ of the sphere, and then the patch anchored to it is described by the coordinates $(T,r,\Omega_{d-2})$ through
\begin{align}
\sinh(t) = \sqrt{1 - r^2}\,\sinh (T)\,,\quad \cosh(t)\,\cos(\theta)  = \sqrt{1 - r^2}\,\cosh(T)\,,\quad \cosh(t)\,\sin(\theta) = r\,.
\end{align}
In these coordinates the line element is
\begin{align}
\label{E:static}
ds^2 = -(1 - r^2)\,dT^2 + \frac{dr^2}{1 - r^2} + r^2\,d\Omega_{d-2}^2\,,
\end{align}
and the generator of time translations is $K = -i \partial_T$. The cosmological horizon is at $r=1$. It has a classical Bekenstein-Hawking entropy $S_0 = \frac{A_{d-2}}{4G}$ with $A_{d-2}=\text{vol}(\mathbb{S}^{d-2})$. Analytically continuing to imaginary time $T = -i\,T_E$ and letting $r = \sin(\psi)$,~\eqref{E:static} locally becomes the round metric on $\mathbb{S}^d$
\begin{align}
ds^2 = d\psi^2 + \sin^2(\psi)\,dT_E^2 + \cos^2(\psi)\,d\Omega_{d-2}^2\,,
\end{align}
which is smooth provided $T_E \sim T_E + 2\pi$, i.e.~the horizon has an inverse Hawking temperature $\beta_{\rm dS} = 2\pi$. For quantum field theory in the static patch we then have
\beq
	\text{tr}\!\left( e^{-\beta_{\rm dS}K}\right) = Z_{\rm matter}\,,
\eeq
with $Z_{\rm matter}$ the sphere partition function and the trace is taken over states in the static patch.\footnote{Although this picture is not quite right since there is no such thing as a reduced density matrix for a local quantum field theory, it is the same style of argument as e.g.~\cite{Casini:2011kv}.} 

\begin{figure}[t!]
\begin{center}
\includegraphics[scale=.4]{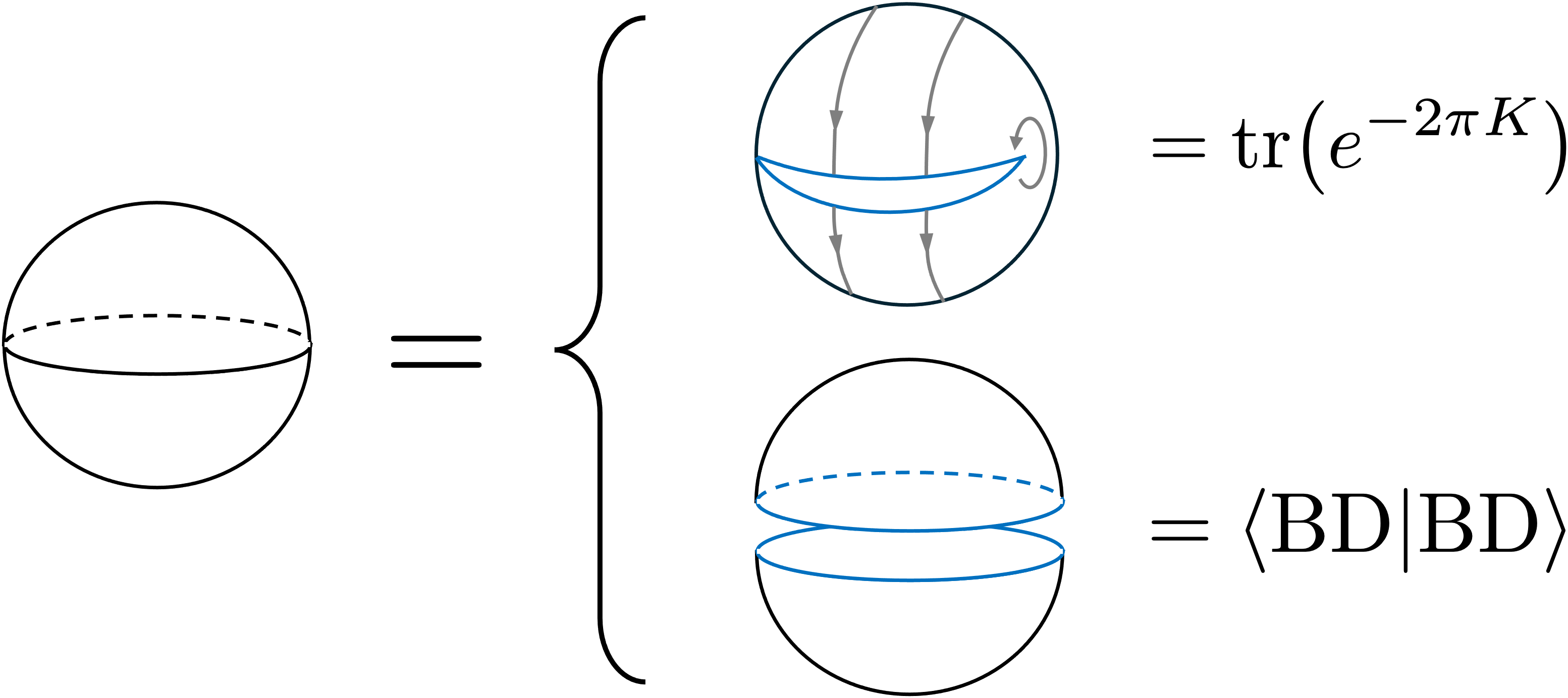}
\end{center}
\caption{The sphere partition function of matter can be understood in two equivalent ways. The first is the thermal partition function for fields in the static patch at inverse temperature $2\pi$, and the second is the norm of the Bunch-Davies state prepared by the hemisphere path integral. \label{Fig:spheretrace1}}
\end{figure}

For a field theory we can cut the sphere along the equator and assign another interpretation to the sphere partition function. The Euclidean path integral over fields on the hemisphere prepares the Bunch-Davies state $|\text{BD}\rangle$, and the sphere amplitude computes the overlap of this state with itself. See Fig.~\ref{Fig:spheretrace1}. We can also take the state on the equator, evolve it in real time (even to the infinite future), and by unitarity the norm remains $Z_{\rm matter}$. 

Gibbons and Hawking proposed that, in a theory of gravity, the sphere partition function $Z_{\rm sphere}$ continues to be a thermal partition function for fields in the static patch. Their basic observation was that, since there is no well-defined notion of gravitational energy in the absence of a boundary on which to anchor, the thermodynamic first law suggests that thermal partition function is merely the exponential of an entropy. Since the tree-level approximation to $Z_{\rm sphere}$ is
\beq
	\ln Z_{\rm sphere} \approx S_0 \,,
\eeq
with $S_0$ the classical horizon entropy, they suggested that $\ln Z_{\rm sphere} $ defines the horizon entropy more generally. If we were so bold as to think of $Z_{\rm sphere}$ as a Hilbert space trace as for fields in the static patch, then effectively $K = 0$ in a theory of gravity, suggesting that $Z_{\rm sphere}$ counts the dimension of the Hilbert space of states in the static patch. (This interpretation is consistent with the gravitational first law whereby the entropy of the static patch decreases when placing a black hole at its center~\cite{Susskind:2021dfc}.)

However these proposals cannot be correct in general $d$ on account of the fact that the renormalized sphere partition function of gravity has a phase as we presently review.

\subsection{Polchinski's phase}
\label{S:phase}

Let us review the one-loop phase of the sphere amplitude of Einstein gravity with positive cosmological constant~\cite{Polchinski:1988ua} described by
\beq
	S_E = - \frac{1}{16\pi G}\int d^dx \sqrt{g}(R - 2\Lambda)\,, \qquad \Lambda = \frac{(d-1)(d-2)}{2}\,.
\eeq
The round metric $g=g_0$ is a saddle point with action $-S_0$. Parameterizing metric fluctuations by $g_{\mu \nu} = g_{0\mu\nu} + \sqrt{32\pi G}\,h_{\mu \nu}$, we decompose $h_{\mu\nu}$ as (see e.g.~\cite{Bastianelli:2013tsa})
\begin{align}
h_{\mu \nu} = \overline{h}_{\mu \nu} + \,g_{\mu \nu} h\,,
\end{align}
with $g_0^{\mu\nu}\overline{h}_{\mu\nu} = 0$. It is convenient to use an $R_\xi$ version of de Donder gauge. In this gauge we add
\beq
	S_{\text{gauge}} =-\int d^{d} x\,\sqrt{g} \,\frac{1}{2}\!\left(\nabla^\nu h_{\nu \mu} -\frac{1}{2d}\nabla_\mu h \right)^{\! 2}
\eeq 
 to the action along with a $bc$ ghost system described by
\beq
S_{bc} = - \int d^{d} x \sqrt{g}\,b^\mu [\nabla^2 c_\mu + R_{\mu \nu} c^\nu] = - \int d^dx \sqrt{g_0} \,b^{\mu}(\nabla^2 +d-1)c_{\mu}\,,
\eeq
where in the second expression we have restricted to quadratic order in fluctuations and $\nabla_{\mu}$ is the covariant derivative on the round sphere. The quadratic action for the fluctuations is
\beq
S_{\rm quad}  =\int d^{d} x \sqrt{g_0} \left(- \frac{1}{2} \overline{h}^{\mu \nu} \!\left( \nabla^2 - 2\right)\overline{h}_{\mu \nu} + \frac{d(d-2)}{4}\,h \left( \nabla^2 +2(d-1)\right)h \right) + S_{\rm bc}\,.
\eeq
Furthermore the Faddeev-Popov procedure instructs us that we should take the absolute value of ghost determinant. The advantage of this gauge choice is that the fluctuations decouple.

The conformal mode is mostly wrong-sign. Under the assumption that the integration contour of the theory can be chosen to run through the sphere saddle along a steepest descent contour we continue $h \to i h$~\cite{Gibbons:1978ac}. Starting from the usual ultralocal measure over metrics this incurs no phase. However, the rotated mode $h$ is now only mostly right-sign. To see this, note that the relevant differential operator has a spectrum
\beq
	-(\nabla^2 + 2(d-1)) \phi = \lambda \phi\,, \qquad \lambda = j(j+d-1) - 2(d-1)\,,
\eeq
equal to that of a tachyonic scalar with mass-squared $M^2 = -2(d-1)$. Here $j$ is the angular momentum on the sphere. The $j>1$ modes have positive eigenvalues, while those with $j=0,1$ are negative. Since there is a single $j=0$ mode and $d+1$ modes with $j=1$, rotating these modes produces a phase $(\pm i)^{d+2}$ depending on whether we rotate by $\pm 90^{\circ}$, although e.g.~\cite{Maldacena:2024spf} has claimed that the correct choice is $i^{d+2}$. All of the other bosonic eigenvalues are positive, while the absolute value of the ghost determinant precludes a phase from the ghost sector. There are ghost zero modes signaling the existence of a residual gauge symmetry, the $SO(d+1)$ isometry of the sphere. In the correct treatment of the ghost sector we omit the path integral over those modes and instead divide by the properly normalized volume of the residual gauge symmetry. This leads to the one-loop result
\beq
\label{E:Poli}
	Z_{\rm sphere} = i^{d+2} e^{S_0} \frac{Z S_0^{-\mathcal{D}_d}}{\text{vol}(SO(d+1))}\,,
\eeq
where $Z$ is a certain ratio of determinants whose renormalized value is finite and non-negative and $\mathcal{D}_d = \text{dim}(SO(d+1)) = \frac{d(d+1)}{2}$. The zeta-regularized version of $Z$ is derived in~\cite{Anninos:2020hfj}.

\section{Late-time overlaps}
\label{S:latetime}

We turn our attention to Einstein gravity with positive cosmological constant at late Lorentzian time in a context where future infinity is a sphere. The main goal of this Section is to establish the expression~\eqref{E:main} for the one-loop approximation to the norm of a state produced by the gravitational path integral, and to flesh out the argument we presented in the Introduction that the norm of the no-boundary state is non-negative, but vanishes at one loop. Along the way we study the enlightening case of pure gravity in $d=3$.

\subsection{Asymptotic states, and the conformal mode}

Ignoring matter for the moment, an asymptotically de Sitter spacetime has a late-time line element that can be put into the form
\beq
\label{E:FG}
	ds^2 = -dt^2 +e^{2t}  \left( \gamma_{ij}(x)+ \cdots\right)dx^i dx^j 
\eeq
as $t\to\infty$, where $\gamma_{ij}$ is held fixed as a future boundary condition and the dots indicate subleading terms whose precise form depends on the curvature of $\gamma$ and the dimension $d$. Here we are choosing a gauge analogous to the Fefferman-Graham gauge for asymptotically locally anti-de Sitter spacetimes. For gravity coupled to matter we ought to also supply late-time boundary conditions on the matter fields, for which we hold the late-time matter profile $\varphi$ fixed.

We interpret the real-time gravitational path integral with $\gamma$ held fixed as an overlap $\langle \!\langle \gamma| \Psi\rangle \! \rangle $ where $|\Psi\rangle \! \rangle$ is the late-time state and the boundary condition of fixed $\gamma$ and matter prepares a bra which we notate as $\langle\! \langle \gamma|$. We term the latter asymptotic states.\footnote{Note that these states are labeled by field configurations rather than particle-like excitations as in e.g.~\cite{Marolf:2012kh}.} 

In ordinary quantum field theory there are analogous asymptotic states $|\varphi\rangle$ in which we hold the quantum fields fixed to a profile $\varphi$ on the late time slice. These states comprise a basis with an ultralocal inner product, schematically $\langle \varphi|\varphi'\rangle = \delta[\varphi-\varphi']$. In the presence of gravity we assume that the asymptotic states corresponding to fixed $(\gamma,\varphi)$, namely $|\gamma,\varphi\rangle \! \rangle$, span a basis, at least in perturbation theory in $G$ and where there is a single universe in the final state. We will have more to say about their inner products later.

The metric $\gamma_{ij}$ resembles the boundary metric appearing in asymptotically AdS spacetimes, and indeed many familiar aspects of stress tensor dynamics in AdS gravity have an analogue in dS gravity, albeit for the classical approximation to $-i \log \Psi_{\rm HH}$, i.e.~the on-shell action. For example we expect the wavefunction to not depend (up to anomalies) on $\gamma$ but rather on $\gamma$ modulo diff$\times$Weyl. The more precise version of this statement is the following. 

Wavefunctions $\langle\!\langle \gamma | \Psi \rangle \! \rangle$ have transformation properties that are a consequence of the diffeomorphism-invariance of the gravitational path integral. Letting $\xi$ be an infinitesimal boundary diffeomorphism corresponding to a bulk diffeomorphism ``parallel'' to the late-time boundary, and letting $\sigma$ be a Weyl rescaling corresponding to a late-time reparameterization $t\to t + \sigma(x)$, we have
\beq
	\langle \! \langle \gamma + \pounds_{\xi}\gamma + 2 \sigma \gamma|\Psi\rangle \! \rangle = \exp\left( - \mathcal{S}[\gamma;\xi,\sigma]\right) \langle \! \langle\gamma|\Psi\rangle \! \rangle\,, \qquad
	\mathcal{S}  = \int d^{d-1}\Omega \sqrt{\gamma} \left( \xi^i \mathcal{A}_i + \sigma \mathcal{A}\right),
\eeq
where $\mathcal{S}$ takes the same form as the anomalous variation of a CFT partition function under infinitesimal diffeomorphisms $\xi$ and Weyl rescalings $\sigma$. For Einstein gravity the diffeomorphism part $\mathcal{A}_i$ vanishes while the Weyl part $\mathcal{A}$ vanishes in even but not odd $d$, taking the same form as the usual Weyl anomaly of a $d-1$ dimensional CFT. 

At tree level and in odd $d$ the anomalous variation $\mathcal{A}$ can be obtained by a continuation of the familiar Weyl anomaly obtained from classical Einstein gravity with negative cosmological constant~\cite{Henningson:1998gx} under $L_{\rm AdS}\to i L_{\rm dS}$~\cite{Maldacena:2002vr} with the result that it is pure imaginary, so that at tree level the unnormalized distribution $|\Psi|^2$ is Weyl-invariant.

Weyl anomalies in CFT are tightly constrained by algebraic consistency. They always have the property that the linearized anomaly action $\mathcal{S}$ vanishes under global conformal transformations, which in CFT implies that global conformal transformations are genuine symmetries even when Weyl-invariance is broken by anomalies. As a result wavefunctions computed from the gravitational path integral are invariant under the action of global conformal transformations on the final state. This will be important later.

In even $d$ we are left with the statement that physical asymptotic states are equivalence classes of $|\gamma\rangle\!\rangle$'s modulo diff$\times$Weyl; that is, the completeness relation for the basis of asymptotic states is $\propto \int \frac{[d\gamma]}{\text{diff}\times\text{Weyl}}|\gamma\rangle\!\rangle\langle\!\langle \gamma|$. In odd $d$ we expect this remains true, namely that the integrand $|\gamma\rangle\!\rangle\langle\!\langle \gamma|$ and integration measure carry equal and opposite Weyl anomalies. This is true to one-loop order in $d=3$ and holds at tree-level in $d>3$ on account of $|\Psi|^2$ being Weyl-invariant at that order.

The Fefferman-Graham-like gauge~\eqref{E:FG} is useful for identifying the asymptotic states $|\gamma\rangle\!\rangle$, however if we are interested in computing loop corrections it is more convenient to use the $R_{\xi}$ version of de Donder gauge used in Subsection~\ref{S:phase} in which the various metric and ghost fluctuations decouple at one-loop order. Recall the discussion there. Metric fluctuations $h_{\mu\nu}$ can be decomposed into a traceful part $h$ and a traceless part $\overline{h}_{\mu\nu}$. The late-time falloffs of these fields can be found by solving their equations of motion $(\nabla^2 - 2)\bar{h}_{\mu\nu}=0$ and $(\nabla^2+2(d-1))h = 0$. For example,
\beq
	h(t,\Omega) = e^{-\Delta_- t}\left( \mathcal{H}_-(\Omega) + \cdots\right)+ e^{-\Delta_+ t}\left( \mathcal{H}_+(\Omega)+ \cdots\right) \,, \quad \Delta_{\pm} = \frac{d-1\pm \sqrt{(d+7)(d-1)}}{2}\,,
\eeq
and we have a consistent variational principle holding the leading growth $\mathcal{H}_-$ fixed as $t\to\infty$. 

Na\"{i}vely the wavefunction depends on $\mathcal{H}_-$ as well as the rest of the leading falloffs in the traceless sector. However this statement is incorrect. The culprit is that there are more solutions to the equations of motion of the gauge-fixed theory than of the original Einstein's equations. To ensure that we obey the same boundary conditions as before we must set $\mathcal{H}_-$ to vanish along with certain leading falloffs for $\overline{h}_{\mu\nu}$. 

A general traceless fluctuation $\overline{h}_{\mu\nu}$ can be decomposed into transverse traceless (TT) modes $h^{TT}$, longitudinal traceless (LT) modes parameterized by a vector $V^T$, and purely longitudinal (LL) fluctuations parameterized by a scalar $\Phi$
\beq
	\overline{h}_{\mu\nu} =h^{TT}_{\mu\nu} + \nabla_{\mu}V^T_{\nu} + \nabla_{\nu}V^T_{\mu} + \left( \nabla_{\mu}\nabla_{\nu} - \frac{1}{d}g_{\mu\nu}\nabla^2\right)\Phi\,,
\eeq
with
\beq
	\nabla^{\nu}h^{TT}_{\mu\nu} = 0\,, \qquad \nabla^{\mu}V^T_{\mu}=0\,.
\eeq
This decomposition is useful as these modes decouple from each other in solving the eigenvalue problem $(\nabla^2 - 2)\phi_{\mu\nu} = \lambda \phi_{\mu\nu}$ and therefore in the computation of the one-loop determinant. In solving for the late-time falloffs we go on-shell, imposing $(\nabla^2 - 2)\overline{h}_{\mu\nu} = 0$, in which case it turns out that this decomposition is overcomplete with the LT modes actually obeying the transverse condition. The linearized solutions to the Einstein's equations then correspond to the set of TT modes modulo the LT subspace. The gauge-fixing term and its variation vanishes on the TT modes implying that the TT modes solve the Einstein's equations, and removing the LT modes simply removes gauge variations. This result also satisfies a simple counting exercise. The number of degrees of freedom in the quotient $\faktor{\text{TT}}{\text{LT}}$ is
\beq
	\underbrace{\left( \frac{d(d+1)}{2}-(d+1)\right)}_{\text{TT}} - \underbrace{(d-1)}_{\text{LT}} = \frac{d(d-3)}{2}\,,
\eeq
which matches both the number of propagating degrees of freedom to the Einstein's equations as well as the number of degrees of freedom in the boundary metric $\gamma$ modulo diff$\times$Weyl.

One punchline is that in de Donder gauge the wavefunction does not actually depend on the conformal mode, which instead contributes to its normalization but does not label physical states at future infinity.

Ultimately this means that to verify that the norm of the no-boundary state is non-negative we must check two things. First, that the inner product of asymptotic states is non-negative, and second that the wavefunction of the no-boundary state is a right-sign Gaussian distribution in the fluctuations of $\gamma$ modulo diff$\times$Weyl, with no zero modes, or equivalently in the $\faktor{\text{TT}}{\text{LT}}$ directions.

\subsection{3d gravity}

\begin{figure}[t!]
\begin{center}
\includegraphics[scale=.4]{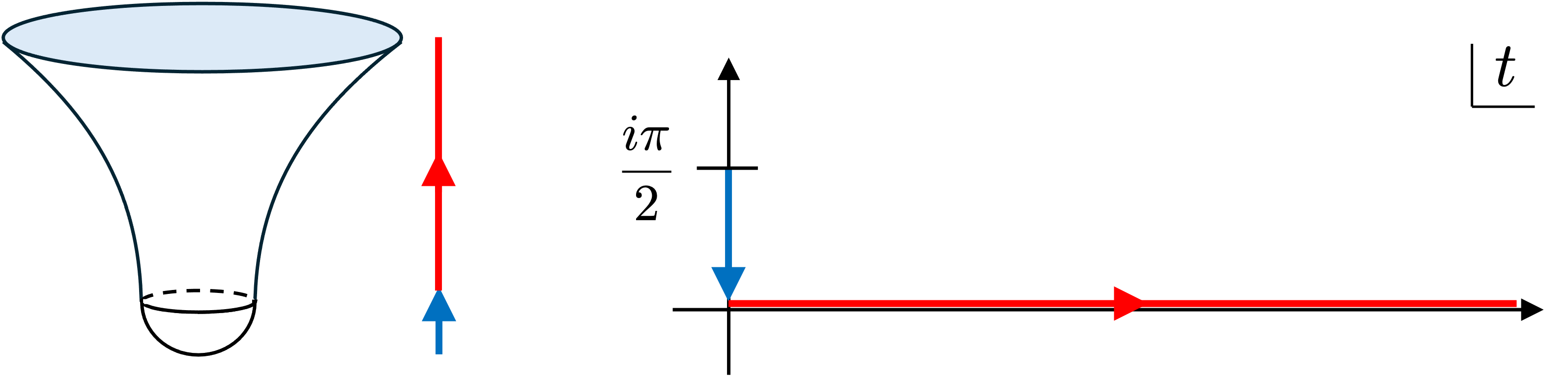}
\end{center}
\caption{The saddle point of the gravitational path integral that computes the late-time no-boundary wavefunction is a complex geometry (left), where time traverses a complex time contour (right). There is a Euclidean time segment (blue) from $\frac{i\pi}{2}$ down to zero, followed by a Lorentzian segment from zero to infinity. \label{Fig:HH}}
\end{figure}

Before discussing the problem of de Sitter gravity in general $d$ we find it instructive to first study the case of pure gravity in $d=3$ in some detail. The simplifying feature is that the Hilbert space of asymptotic states is one-dimensional, since the metric $\gamma$ on the late-time sphere can be completely fixed to the round one by diff$\times$Weyl. To begin we must set the stage by reviewing the semiclassical approximation to the wavefunction of the no-boundary state.

The late-time wavefunction of the Bunch-Davies state is computed by a path integral with a complex-time trajectory in which a Euclidean hemisphere prepares the initial state which is then evolved in real Lorentzian time. The spacetime is described by the line element
\beq
\label{E:globaldS}
	ds^2 = -dt^2 + \cosh^2(t)d\Omega_{d-1}^2\,,
\eeq
where $t$ traverses the trajectory indicated in Fig.~\ref{Fig:HH}, first going as $t = i\tau$ from $\tau=\pi/2$ to $\tau=0$, and then evolving from $t=0$ to $t\to\infty$ with the late-time boundary conditions specified there. The real-time segment of the contour is the future half of global de Sitter space, and the gluing to the hemisphere along $t=0$ is smooth on account of the fact that the slice $t=0$ has zero extrinsic curvature. With perturbative gravity we work in the same way, assuming that the late-time wavefunction of the no-boundary state is computed by the gravitational path integral over fields on such a complex time contour. Effectively we are summing over (complex) geometries that smoothly fill in the late-time boundary condition, where the spacetime~\eqref{E:globaldS} and complex time trajectory give the dominant saddle for that problem when the future boundary condition is that $\gamma$ is the round metric on the sphere. We call this spacetime (both when $\gamma$ is the round metric and also when it is not) the Hartle-Hawking geometry.

In $d=3$ we may use the diff$\times$Weyl symmetry at future infinity to fix $\gamma$ to be exactly the round metric $\gamma_0$. With that choice the wavefunction is simply a number,
\beq
	\Psi_{\rm HH} = \langle\!\langle \gamma_0 |\text{HH}\rangle\!\rangle = \Psi_{1\text{-loop}}\,e^{\frac{S_0}{2}}\,,
\eeq
where we denote the late-time no-boundary state as $|\text{HH}\rangle\!\rangle$. It may be computed to one loop by integrating over fluctuations around the Hartle-Hawking geometry that prepares the late-time state, with on-shell action $iS = \frac{S_0}{2}$ and $\Psi_{1\text{-loop}}$ a certain ratio of one-loop determinants.\footnote{It is easy to check that the fluctuation spectrum is a continuum of ``scattering states'' of nonzero modes. In fact this continuum is responsible for the $+13$ shift in the effective central charge.} Under Weyl rescalings of $\gamma_0$ the wavefunction transforms as a CFT partition function with an effective central charge $c = \frac{3i}{2G}+13$ to one loop. The leading term comes from the classical gravity approximation and the $13$ is a one-loop correction~\cite{Cotler:2019nbi}. Like the holographic Weyl anomaly, the effective central charge comes from the infinite spacetime volume near future infinity. The unnormalized distribution $|\Psi_{\rm HH}|^2$ is also anomalous with effective central charge $c + c^* = 26$. 

\begin{figure}[t!]
\begin{center}
\includegraphics[scale=.45]{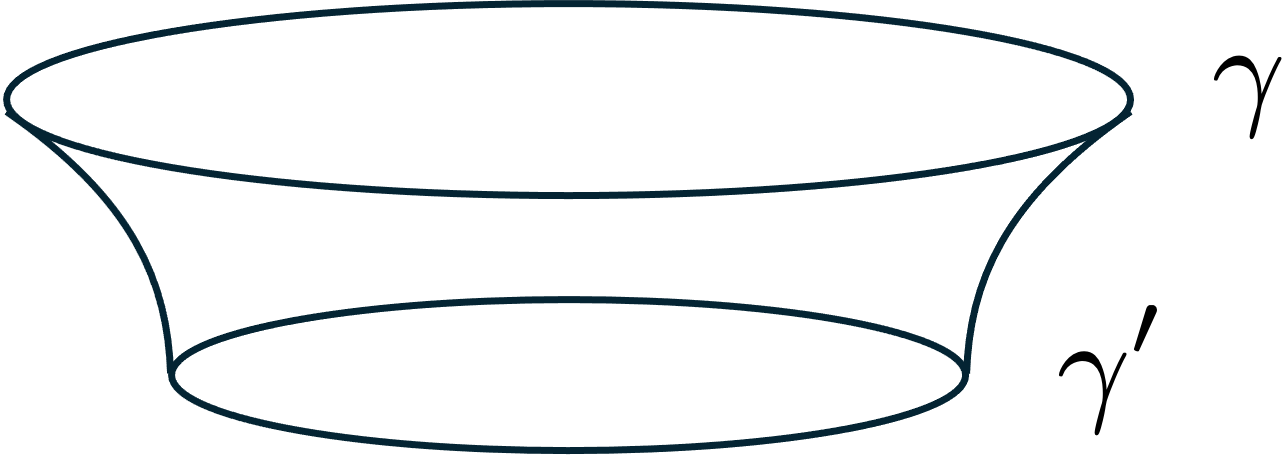}
\end{center}
\caption{The inner product $\langle\!\langle \gamma|\gamma'\rangle\!\rangle$ between asymptotically de Sitter states corresponds to computing the transition between two late-time slices in the asymptotic future $t \to \infty$. \label{Fig:IP1}}
\end{figure}

To make sense of the norm we require the overlap of the state $|\gamma_0\rangle\!\rangle$ with itself, so that we can construct an identity operator $\frac{|\gamma_0\rangle\!\rangle\langle\!\langle \gamma_0|}{\langle \!\langle \gamma_0|\gamma_0\rangle\!\rangle}$ on this Hilbert space. We contend that the gravitational path integral can be used to compute that inner product through a short-time gravitational path integral over short cylinders that connect two asymptotic boundaries, one that prepares an asymptotic ket and the other an asymptotic bra. See Fig.~\ref{Fig:IP1} for a depiction when the metric $\gamma'$ labeling the ket differs from that $\gamma$ labelling the bra. In de Sitter JT gravity (see e.g.~\cite{Maldacena:2019cbz, Cotler:2019nbi, Moitra:2021uiv, Held:2024rmg}) we can compute this inner product~\cite{Cotler:2024xzz} (building on~\cite{Cotler:2019dcj, Cotler:2023eza}) but in Einstein gravity it is more involved~\cite{CJWIP1}. 

We claim that the inner product $\langle\!\langle \gamma|\gamma\rangle\!\rangle$ satisfies two properties. The first is that under Weyl rescalings it transforms with effective charge $c+c^* = 26$, i.e.~in the same way as the outer product $|\gamma\rangle\!\rangle\langle\!\langle \gamma|$ so that the identity operator $\frac{|\gamma\rangle\!\rangle\langle\!\langle \gamma|}{\langle \!\langle \gamma|\gamma\rangle\!\rangle}$ is Weyl-invariant. The second is that at one-loop it is equal to a non-negative renormalized constant times the volume of the conformal group on the sphere, $\text{vol}(SO(3,1))$. 

While a complete computation of the inner product will be done elsewhere~\cite{CJWIP1}, we can argue that the above claims are true using the global de Sitter amplitude, i.e.~the gravitational path integral over fluctuations around empty global dS$_3$ space. We interpret that amplitude as the matrix element of the infinite-time evolution operator $\hat{\mathcal{U}}$ from past to future infinity,
\begin{align}
     \includegraphics[scale=.32, valign = c]{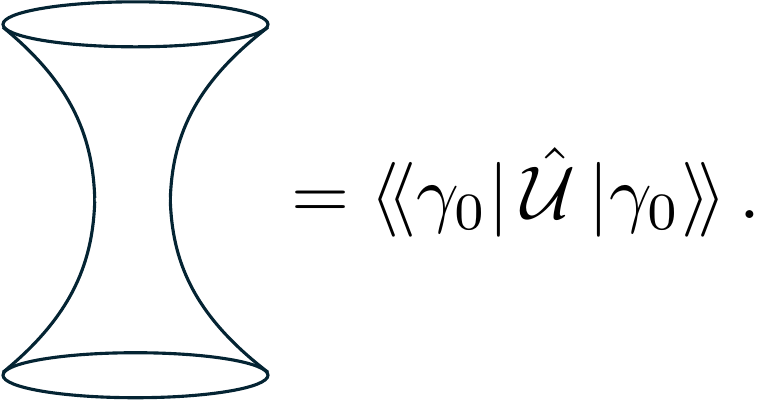}
\end{align}
In the canonical approach to gravity, which we expect to be valid in this semiclassical regime, the Hamiltonian of a closed universe vanishes and so $\hat{\mathcal{U}} $ ought to be the identity, or perhaps some anomalous phase times the identity.  (See~\cite{Blommaert:2025bgd} and also~\cite{Cotler:2024xzz} for recent discussions of this point.) That argument, combined with the assertion that we should define the measure of the gravitational path integral so that the inner product is positive, implies that the inner product is the modulus of this global de Sitter amplitude. Relatedly, another argument for this claim is by contradiction. If it is not true, then there are violations of unitarity of evolution visible at one loop. By this argument the ``long-time'' amplitude $\langle\!\langle \gamma_0|\,\hat{\mathcal{U}}\,|\gamma_0\rangle\!\rangle$ gives the ``short-time'' amplitude $\langle\!\langle \gamma_0|\gamma_0\rangle\!\rangle$. 

In the classical approximation the on-shell action of the global dS$_3$ geometry vanishes. At one loop the spectrum of fluctuations includes a continuum of ``scattering'' states, with a density of states agreeing with that governing $|\Psi_{\rm HH}|^2$, as well as a discretuum of ``bound states.'' The former imply that the long-time amplitude has an effective central charge of $26$ as claimed. Physically, this comes from the fact that there is a future asymptotic region which contributes $c = \frac{3i}{2G}+13$ (see~\cite{Cotler:2019nbi} for a one loop computation), and a past asymptotic region contributing $c^*$.

The volume of the conformal group arises from certain bosonic zero modes of global dS$_3$ which we call ``conformal twists.'' These are large diffeomorphisms that we integrate over in the gravitational path integral, acting as the identity on one boundary (say past infinity) while acting as a conformal transformation on the other. These modes are analogous to the twists of JT gravity, which are exact zero modes of the global de Sitter saddle, although here they are noncompact rather than compact as per JT gravity.

The twists are certain zero modes with one unit of angular momentum on the spatial two-sphere. These modes are both TT and LT and thus can be written as large diffeomorphisms. They obey
\beq
	(\nabla^2 - 2)\overline{h}_{\mu\nu} = 0\,, \qquad \nabla^{\mu}\overline{h}_{\mu\nu} = 0\,, \qquad \overline{h}_{\mu\nu}=\nabla_{\mu}V^T_{\nu}+\nabla_{\nu}V^T_{\mu}\,,\qquad \nabla^{\mu}V^T_{\mu} = 0\,,
\eeq
and are given by
\beq
	\sqrt{32\pi G}\,\overline{h}_{\mu\nu} dx^{\mu}dx^{\nu} =A^a\left( f_1(t)(2 dt^2 + \cosh^2(t) d\Omega^2) X_a + 2 f_2(t) dt \,dX_a\right)+2B^af_4(t)  dt d\theta^{\alpha}\epsilon_{\alpha\beta} \hat{\nabla}^{\beta} X_a\,,
\eeq
where we label the six modes with constants $A^{a=1,2,3}$ and $B^{a=1,2,3}$, parameterize the sphere with $X_{a=1,2,3}$ satisfying $X_aX_a = 1$, $\hat{\nabla}_{\alpha}$ and $\epsilon_{\alpha\beta}$ are the covariant derivative and epsilon tensor on the sphere, sphere indices $\alpha$ are raised and lowered using the metric on the sphere, and
\beq
	f_1  =- \frac{3}{4} \,\text{sech}^4(t)\,, \qquad f_2 = -\frac{3}{4}\, \text{sech}^2(t)\tanh(t)\,, \qquad	f_3 = \frac{3}{4}\,\text{sech}^2(t)\,.
\eeq
These modes can be written as large diffeomorphisms up to integration constants that can be thought of as isometries (which for de Sitter are large and so are not gauged), with 
\beq
	\sqrt{32\pi G}\,V^T_{\mu}dx^{\mu} = A^a\left( h_1(t) dt X_a + h_2(t) dX_a \right)+B^a h_3(t) d\theta^{\alpha}\epsilon_{\alpha\beta} \hat{\nabla}^{\beta} X_a\,,
\eeq
and
\beq
	h_1 = -\frac{1}{4}(2+3\tanh(t) - \tanh^3(t))\,, \qquad h_2 = \frac{1}{8}(2e^{2t} + \text{sech}^2(t) )\,, \qquad h_3 = \frac{e^{2t}}{4}(2-\tanh(t))\,.
\eeq
At large positive and negative time the angular components of $V^{T\mu}$ are
\begin{align}
\begin{split}
	\sqrt{32\pi G}\,V^{T\alpha} & = \begin{cases} A^a  \hat{\nabla}^{\alpha} X_a + B^a \epsilon^{\alpha\beta} \partial_{\beta} X_a  + O(e^{-2|t|})\,, & t \to \infty\,, \\ O(e^{-2|t|}) \,, & t\to -\infty\,.\end{cases}
\end{split}
\end{align}
In the far past these large diffeomorphisms vanish while in the far future they parameterize an infinitesimal conformal transformation. To see this, label
\beq
	\sqrt{32\pi G}\,V^{T\alpha} \partial_{\alpha}= A^a K_a + B^a R_a + O(e^{-2|t|})\,,
\eeq
where the $K_a$ generate boosts and $R_a$ the rotations. Expressing the $X^a$ in spherical coordinates,
\beq
	X^1 = \sin(\theta)\cos(\phi)\,, \qquad X^2 = \sin(\theta)\sin(\phi)\,, \qquad X^3 = \cos(\theta)\,,
\eeq
we have
\begin{align}
\begin{split}
	R_1 &= - \left( \sin(\phi)\partial_{\theta}+\cot(\theta)\cos(\phi)\partial_{\phi}\right) \,, 
	\\
	R_2& = \cos(\phi)\partial_{\theta} -\cot(\theta)\sin(\phi)\partial_{\phi}\,,
	\\
	 R_3 &= \partial_{\phi}\,, 
	\\
	K_1 & = \cos(\theta)\cos(\phi)\partial_{\theta}-\csc(\theta)\sin(\phi)\partial_{\phi}\,, 
	\\
	 K_2 &= \cos(\theta)\sin(\phi)\partial_{\theta}+\csc(\theta)\cos(\phi)\partial_{\phi}\,,
	 \\
	  K_3 &=  -\sin(\theta)\partial_{\theta}\,.
\end{split}
\end{align}
These are the (anti-Hermitian) generators of $SO(3,1)$ whose commutators of Lie derivatives obey
\begin{align}
\begin{split}
	[K_a,K_b] & = \epsilon_{abc}R_c\,,
	\\
	[K_a,R_b] & = - \epsilon_{abc} K_c\,,
	\\
	[R_a,R_b] &= - \epsilon_{abc}R_c\,.
\end{split}
\end{align}
Note that the rotations are normalized so that $\exp(2\pi R)=1$.

The metric on these zero modes comes from the ultralocal inner product
\beq
	(\overline{h},\overline{h}) =- \frac{1}{2\pi}  \int d^3x \sqrt{-g} \,\overline{h}_{\mu\nu}\overline{h}^{\mu\nu}\,,
\eeq
(we introduce a minus sign because the overlap is usually negative in Lorentzian signature) giving
\beq
	ds^2_{\rm PI} = \frac{1}{16\pi G} \delta_{ab}(-A^a A^b +B^a B^b) = \frac{1}{16\pi G} ds^2_c\,,
\eeq
where $ds^2_c$ is the canonically normalized metric on $SO(3,1)$. The properly normalized integral over these zero modes using $S_0 = \frac{\pi}{2G}$ is then
\beq
	\text{vol}(\text{ZM}) \propto S_0^{3} \,\text{vol}(SO(3,1))\,.
\eeq
The powers of $G$ can also be guessed from the fact that there are six bosonic zero modes.

Our expectation is that the inner product is an integral over fluctuations around the global de Sitter cylinder connecting the $t=T$ slice to the $t=T+\delta$ slice in the limit $\delta,T\to \infty$. The same continuum of modes that contribute to the global dS amplitude contribute here, producing an effective central charge of 26, and we can rescale the twists of the long-time amplitude to act as conformal twists in this problem. By adding an isometry to $V^T_{\mu}$ we can tune these short-time twists to act as the identity on the $t=T$ slice and as a conformal transformation on the $t=T+\delta$ slice. 

In the bare-bones de Donder gauge there is the troublesome feature that there are residual ``small'' gauge transformations, leading to zero modes in the LT, TT, and gauge sectors. These zero modes are unphysical, an artifact of an incomplete specification of gauge, and after fixing it only nonzero modes remain. As a result the long-time amplitude is
\beq
	\langle \!\langle \gamma_0|\,\hat{\mathcal{U}}\,|\gamma_0\rangle\!\rangle = \mathcal{Z} S_0^3 \text{vol}(SO(3,1))\,,
\eeq
for $\mathcal{Z}$ a ratio of renormalized one-loop determinants over nonzero modes whose precise value is irrelevant in what follows. By our argument above that the inner product is the modulus of this amplitude, the contribution of one-sphere states to the norm of the no-boundary state is
\beq
	\boxed{ \langle \!\langle \text{HH}| \text{HH}\rangle\!\rangle =\frac{\langle \!\langle\text{HH}|\gamma_0\rangle\!\rangle\langle\!\langle \gamma_0|\text{HH}\rangle\!\rangle}{\langle\!\langle \gamma_0|\gamma_0\rangle\!\rangle}  +\cdots = e^{S_0} \frac{\widetilde{Z} S_0^{-3}}{\text{vol}(SO(3,1))} +\cdots}
\eeq
where $\widetilde{Z} = |\Psi_{1\text{-loop}}|^2/ |\mathcal{Z}|$ and the dots indicate higher loops and other topologies. This result is the simplest realization of the various claims advertised in the Introduction. In particular, the sphere contribution to the norm is a non-negative constant divided by infinity.

This result is consistent with our main claim~\eqref{E:main} that the contribution of one-sphere states to the norm of a state is given by (specializing to $d=3$ and pure gravity) 
\begin{align}
\begin{split}
\label{E:3dstring}
	\langle\!\langle \text{HH}|\text{HH}\rangle\!\rangle &=  \int \frac{[d\gamma]}{\text{diff}\times\text{Weyl}}|\Psi_{\rm HH}[\gamma]|^2 \\
	&= A S_0^{-3} \int \frac{[d\gamma]}{SO(3,1)} \delta[\gamma-\gamma_0] |\text{det}'(K_{\text{ghost}})| |\Psi_{\rm HH}|^2
	\\
	& = e^{S_0} \frac{\widetilde{Z}S_0^{-3}}{\text{vol}(SO(3,1))} \,,
\end{split}
\end{align}
where $K_{\text{ghost}}$ denotes the ghost kinetic operator, $A$ denotes the normalization of the measure together with the factors of $S_0$ chosen to match the 3d gravity result, and $\widetilde{Z} = A |\Psi_{1\text{-loop}}|^2 |\text{det}'(K_{\text{ghost}})|$.  The above can be used to infer $A^{-1} = |\mathcal{Z} \text{det}'(K_{\text{ghost}})|$. Cast this way, we see that the late-time norm is an emergent bosonic string theory, with the vanishing of the sphere contribution to the norm mirroring the usual vanishing of the sphere amplitude of the bosonic critical string with compact target.

While we have computed the one-loop approximation to the norm here, we expect the sphere contribution to vanish to all loops. After all, higher loops can change the constant $\Psi_{\rm HH}$ and rescale the inner product $\langle \!\langle \gamma|\gamma\rangle\!\rangle$, but they will not lift the zero modes found above. Equivalently, from the point of view of an integral over metrics mod diff$\times$Weyl, the residual $SO(3,1)$ symmetry persists to all loops. As a result the leading contribution to the norm of the no-boundary state comes from geometries where future infinity has genus $g\geq 1$ as studied in~\cite{Godet:2024ich,Collier:2025lux}.

So far in this Section we have focused on the case of pure 3d gravity. It is easy to consider instead the problem of gravity coupled to a free scalar field $\phi$ of mass-squared $M^2$ for sufficiently light $M^2$. The late-time falloff of $\phi$ is
\beq
\label{E:phifalloff1}
	\phi(t,\Omega) = e^{-\Delta_-t}\left( \varphi(\Omega)+\hdots\right) + e^{-\Delta_+ t} \left(\zeta(\Omega) + \hdots\right)\,, \quad \Delta_{\pm} = \frac{d-1}{2} \pm \sqrt{\left( \frac{d-1}{2}\right)^2 -M^2}\,,
\eeq
and we consider $0\leq |M| < \frac{d-1}{2}$ so that $\Delta_{\pm}$ are real with $\Delta_-<\Delta_+$. Then we fix the metric $\gamma$ and $\varphi$ at future infinity producing asymptotic states $\langle \!\langle \gamma,\varphi|$ and under Weyl rescalings $t\to t+\sigma(\Omega)$ we have $\gamma_{ij} \to e^{2\sigma}\gamma_{ij}$ and $\varphi\to e^{-\Delta_-\sigma}\varphi$. We can continue to use the diff$\times$Weyl symmetry to fix $\gamma$ to the round metric, leaving a residual $SO(3,1)$ gauge symmetry which acts on the scalar profile. The asymptotic states are the tensor product of a one-dimensional Hilbert space for the metric with the states labeled by fixed $\varphi$, and the Hilbert space is this space subject to an $SO(3,1)$ identification. 

With this gauge choice the no-boundary wavefunction factorizes to one loop
\beq
	\Psi[\varphi]= \langle\!\langle \gamma_0,\varphi|\text{HH}\rangle\!\rangle= \Psi_{\rm HH} \Psi_{\rm BD}[\varphi] \,.
\eeq
In the absence of gravity the inner product of scalar states $|\varphi\rangle$ is $\langle \varphi|\varphi'\rangle = \delta[\varphi-\varphi']$, coming from the ultralocal inner product of states of definite $\phi$ on constant time slices. This amplitude is computed by a path integral in which one integrates over fluctuations around a short-time trajectory that interpolates between the initial and final configurations. The weight $e^{iS}$ of the latter is proportional to a delta functional in the short-time limit, which is properly normalized thanks to the integral over fluctuations around it. Including gravitational effects amounts to including the determinant $|\mathcal{Z}|$ over nonzero modes as well as an integral over the conformal twists. However these twists now act on the scalar data, say $\varphi$ if we let the twists hold the past boundary fixed and only act on the future one. The inner product in the presence of gravity then becomes what we call a twirled version of the QFT inner product
\beq
	\langle\!\langle \varphi|\varphi'\rangle\!\rangle = |\mathcal{Z}| S_0^{3} \int_{SO(3,1)}\!\!\! d\alpha\, \langle \varphi|\hat{U}(\alpha)|\varphi'\rangle\,,
\eeq
where $\alpha\in SO(3,1)$, $d\alpha$ is the Haar measure, and $\hat{U}(\alpha)$ enacts the conformal twist.

The reader will notice that this integral is a group-averaged inner product along the lines of~\cite{Higuchi:1991tk, Higuchi:1991tm}. Here we see that the gravitational path integral naturally produces it as the inner product of asymptotic states. Moreover, the twists play an essential role. On general grounds amplitudes produced by the gravitational path integral must be invariant under independent $SO(3,1)$ transformations acting on the bra and on the ket. The twirling enacted by the twists ensures that this is the case.

This inner product is consistent with a resolution of the identity
\beq
	\hat{\mathbb{1}} = \int \frac{[d\gamma][d\varphi]}{\text{diff}\times\text{Weyl}}|\varphi\rangle\!\rangle\langle\!\langle \varphi|\,,
\eeq
in the gauge where $\gamma$ has been fixed to $\gamma_0$. To see that this is the case, note (using the same normalization for the measure that we inferred in~\eqref{E:3dstring})
\begin{align}
\begin{split}
\label{E:inversion}
	\langle\!\langle \varphi|\hat{\mathbb{1}} |\varphi'\rangle\!\rangle & = \frac{S_0^{-3}}{|\mathcal{Z}|} \int \frac{[d\gamma][d\varphi'']}{SO(3,1)} \delta[\gamma-\gamma_0]  \langle \!\langle \varphi|\varphi''\rangle\!\rangle \langle \!\langle \varphi''|\varphi'\rangle\!\rangle
	\\
	& = |\mathcal{Z}|S_0^{3} \int [d\varphi''] \int_{SO(3,1)} \!\!\! d\alpha d\alpha' \frac{\langle \varphi|\hat{U}(\alpha)|\varphi''\rangle\langle \varphi''|\hat{U}(\alpha')|\varphi'\rangle}{\text{vol}(SO(3,1))}
	\\
	& = |\mathcal{Z}|S_0^3 \int_{SO(3,1)}\!\!\! d\beta \langle \varphi|\hat{U}(\beta)|\varphi'\rangle = \langle\!\langle \varphi|\varphi'\rangle\!\rangle\,,
\end{split}
\end{align}
where in going from the second line to the third we redefined the $\varphi''$ eigenstates as $\hat{U}(\alpha)|\varphi''\rangle = |\varphi'''\rangle$ (which works away from configurations of $\varphi$ of measure 0), traded $\alpha'$ for $\beta = \alpha^{-1}\cdot \alpha'$, and used that the measure for $\varphi''$ and $\beta$ transform with unit Jacobian under these changes. The integral over $\alpha$ becomes trivial and produces a factor of $\text{vol}(SO(3,1))$ canceling against the denominator.

Now including matter the one-sphere contribution to the norm of the no-boundary state is
\begin{align}
\begin{split}
	\langle\!\langle \text{HH}|\text{HH}\rangle\!\rangle &= AS_0^{-3} \int \frac{[d\gamma][d\varphi]}{\text{diff}\times\text{Weyl}}|\Psi[\gamma,\varphi]|^2 =  e^{S_0} \widetilde{Z}S_0^{-3}\times \int \frac{[d\varphi]}{SO(3,1)}|\Psi_{\rm BD}[\varphi]|^2 
	\\
	& =   e^{S_0}\frac{ \widetilde{Z}S_0^{-3}}{\text{vol}(SO(3,1))}\langle \text{BD}|\text{BD}\rangle\,.
\end{split}
\end{align}
The first equality is a more sophisticated example of our general result~\eqref{E:main}. To obtain the second line we used that the Bunch-Davies wavefunction is conformally invariant so that the division by the residual conformal isometry $SO(3,1)$ continues to produce a simple factor of its inverse volume. Note that despite being conformally invariant, $\langle \text{BD}|\text{BD}\rangle$ equals the sphere partition function of the matter, a finite quantity upon renormalization, so that again the one-sphere contribution to the norm vanishes.

\subsection{Overlaps in $d>3$ at one loop}
\label{subsec:overlaps1}

In the last Subsection we established our main result~\eqref{E:main} for the norm of a late-time state for pure 3d gravity and for 3d gravity coupled to matter. Now we argue for~\eqref{E:main} in $d>3$ dimensions. The argument is structurally identical to that appearing in our discussion of 3d gravity coupled to matter, only now we also have to contend with the physical fluctuations of the boundary metric.

Let us consider pure gravity for simplicity, and study asymptotic states where the boundary metric is a small fluctuation away from the round one, $\gamma = \gamma_0 + \sqrt{G}\,\mathfrak{h}$. We fix a gauge whereby $\mathfrak{h}$ is transverse and traceless. Let us label these asymptotic states as $|\mathfrak{h}\rangle\!\rangle$. The one-loop and small $\mathfrak{h}, \mathfrak{h}'$ approximation to the global dS amplitude $\langle\!\langle \mathfrak{h}|\,\hat{\mathcal{U}}\,|\mathfrak{h}'\rangle\!\rangle$ takes the form of a one-loop determinant times the exponential of $iS$ with $S$ the on-shell trajectory interpolating between $\mathfrak{h}'$ and $\mathfrak{h}$. The fluctuations contributing the one-loop determinant are those around empty global dS$_d$. These include conformal twists as in 3d, namely large diffeomorphisms corresponding to normalizable metric fluctuations which can be taken to act as the identity in the past and as conformal transformations at future infinity. The twists are labeled by $A^a$ and $B^{ab}=-B^{ba}$ with $a=1,...,d$ and take the form
\beq
	\sqrt{32\pi G}\,\overline{h}_{\mu\nu} dx^{\mu}dx^{\nu} = A^a \left( f_1(t) ((d-1)dt^2 + \cosh^2(t)d\Omega_{d-1}^2) X_a + 2f_2(t) dt dX_a\right) + B^{ab} f_3(t)dt d\theta^{\alpha} \mathcal{M}_{\alpha ab}\,,
\eeq
with $X_aX_a = 1$ parameterizing the $\mathbb{S}^{d-1}$, and where
\beq
	\mathcal{M}_{\alpha ab} = \frac{1 }{(d-2)!}\epsilon_{abc_1\hdots c_{d-2}} \epsilon_{\alpha}{}^{ \beta_1\hdots \beta_{d-2}}\partial_{\beta_1}X_{c_1} \hdots \partial_{\beta_{d-2}}X_{c_{d-2}}\,,
\eeq
with $\epsilon_{a_1\hdots a_d}$ the flat space epsilon tensor and $\epsilon^{\alpha_1\hdots \alpha_{d-1}}$ the epsilon tensor on the $\mathbb{S}^{d-1}$. Then
\begin{align}
\begin{split}
	f_1(t) &=- \frac{2\Gamma\left(\frac{d+2}{2}\right)}{\sqrt{\pi}(d-1)\Gamma\left(\frac{d+1}{2}\right)} \,\text{sech}^{d+1}(t)\,, 
	\\
	f_2(t) &=- \frac{2\Gamma\left(\frac{d+2}{2}\right)}{\sqrt{\pi}(d-1)\Gamma\left(\frac{d+1}{2}\right)} \,\text{sech}^{d-1}(t)\tanh(t)\,.
	\\
	f_3(t) & =  \frac{\Gamma\left(\frac{d+2}{2}\right)}{\sqrt{\pi}\Gamma\left(\frac{d+1}{2}\right)}\, \text{sech}^{d-1}(t)\,.
\end{split}
\end{align}
These can be written as large diffeomorphisms $\overline{h}_{\mu\nu} = \nabla_{\mu}V^T_{\nu}+\nabla_{\nu}V^T_{\mu}$, which after adding a suitable isometry become
\beq
	\sqrt{32\pi G}\,V^T_{\mu} dx^{\mu}= A^a \left( h_1(t) dt X_a + h_2(t) dX_a \right) + \frac{1}{2}B^{ab}h_3(t)\mathcal{M}_{\alpha ab} \,, 
\eeq
with
\begin{align}
\begin{split}
	h_1(t)  &= -\frac{\Gamma\left(\frac{d+2}{2}\right)}{\sqrt{\pi}\Gamma\left(\frac{d+1}{2}\right)}\,\int_{-\infty}^t dt' \,\text{sech}^{d+1}(t') \,,
	\\
	h_2(t) &= \frac{1}{2}\left(\frac{2\Gamma\left(\frac{d+2}{2}\right)}{\sqrt{\pi}(d-1)\Gamma\left(\frac{d+1}{2}\right)} \text{sech}^{d-1}(t) - \sinh(2t) h_1(t)\right)\,,
	\\
	h_3(t) & =  \frac{\Gamma\left(\frac{d+2}{2}\right)}{\sqrt{\pi}\Gamma\left(\frac{d+1}{2}\right)}\,\cosh^2(t)\int_{-\infty}^t dt' \,\text{sech}^{d+1}(t')\,.
\end{split}
\end{align}
At late positive and negative time the angular components of $V^{T\mu}$ are, thanks to $\int_{-\infty}^{\infty}dt\,\text{sech}^{d+1}(t) = \frac{\sqrt{\pi}\Gamma\left( \frac{d+1}{2}\right)}{\Gamma\left( \frac{d+2}{2}\right)}$,
\beq
	\sqrt{32\pi G}\,V^{T\alpha} = \begin{cases} A^a \hat{\nabla}^{\alpha}X_a +\frac{1}{2}B^{ab} \mathcal{M}^{\alpha}{}_{ab} + O(e^{-2|t|})\,, & t\to\infty\,, \\ O(e^{-2|t|})\,, & t \to -\infty\,.\end{cases}
\eeq
Here the angular index of $\mathcal{M}$ has been raised using the metric on the $\mathbb{S}^{d-1}$. This vector field dies off in the far past and acts as an infinitesimal conformal transformation in the far future. Defining the vector fields $K_a$ and $M_{ab}$ through
\beq
	\sqrt{32\pi G}\,V^{T\alpha}\partial_{\alpha} = A^aK_a + \frac{1}{2}B^{ab}M_{ab}+O(e^{-2t})\,,
\eeq
at late positive time, the $K_a$ generate boosts and $M_{ab}$ rotations with a commutator of Lie derivatives
\begin{align}
\begin{split}
	[K_a,K_b] &= M_{ab}\,,
	\\
	[K_a,M_{bc}] & = \delta_{ab}K_c - \delta_{ac} K_b \,,
	\\
	[M_{ab},M_{cd}] & =\delta_{ac}M_{bd} - \delta_{ad}M_{bc} - \delta_{bc}M_{ad} + \delta_{bd}M_{bc} \,,
\end{split}
\end{align}
reproducing the usual conformal algebra in terms of anti-Hermitian generators. These modes are normalizable with respect to the ultralocal inner product
\beq
	(\overline{h},\overline{h}) = - \frac{1}{2\pi}\int d^dx \sqrt{-g} \,\overline{h}_{\mu\nu}\overline{h}^{\mu\nu}
\eeq	
that we use for the path integral measure. The twist integration measure is
\beq
	\text{vol}_{\rm  PI} \propto S_0^{\mathcal{D}_d/2} \text{vol}_c\,,
\eeq
with $\text{vol}_c$ the canonical integration measure on $SO(d,1)$ in which rotations are $2\pi$ periodic.

The twists act on the boundary data as in our example of 3d gravity coupled to matter. They also appear in the spectrum of fluctuations around a truncated version of the de Sitter cylinder between time $t=T$ and $t=T+\delta$, which we expect describes the inner product in the limit $\delta\to \infty$ and $T\to\infty$. 

Ignoring the twists for the moment, the gravity path integral interpolating between these two asymptotic slices is over metrics that can be decomposed into fluctuations around a trajectory that interpolates between the boundary conditions. The action of the interpolating trajectory is large and after accounting for the nonzero modes we will find a result proportional to a delta function $|\mathcal{Z}|\langle \mathfrak{h}|\mathfrak{h}'\rangle = |\mathcal{Z}|\delta[\mathfrak{h}-\mathfrak{h}']$.  We implicitly choose to define the measure of the gravitational path integral with the one-loop constant $|\mathcal{Z}|$ non-negative so that the inner product is positive-definite. Now restoring the twists, we find at one loop a twirled inner product 
\beq
\label{E:twirledInnerProduct}
	\langle \!\langle \mathfrak{h}|\mathfrak{h}'\rangle\!\rangle = |\mathcal{Z}| S_0^{\mathcal{D}_d/2} \int_{SO(d,1)} \!\!\! d\alpha \,\langle \mathfrak{h}|\hat{U}(\alpha)|\mathfrak{h}'\rangle\,,
\eeq
where $\alpha\in SO(d,1)$ parameterizes the twist, $\hat{U}(\alpha)$ enacts a conformal transformation, and $d\alpha$ refers to the Haar measure over $SO(d,1)$. Also $\mathcal{D}_d = \text{dim}(SO(d,1)) = \frac{d(d+1)}{2}$. The factors of $S_0$ arise from the properly normalized measure over the twists as in our 3d example. 

On general grounds the one-loop inner product produced by the gravitational path integral must be invariant under independent $SO(d,1)$ transformations acting on the bra and on the ket. The argument is that conformal transformations are the $\text{diff}\times\text{Weyl}$ transformations that preserve the round sphere and generate no anomalous variation in odd $d$, and so are residual symmetries of the wavefunction, i.e.~when acting on the ket. But the same argument implies a residual symmetry acting on the bra. In any case the twists implement that property rather minimally through the twirling of~\eqref{E:twirledInnerProduct}. Note that nonzero modes cannot render the one-loop inner product $SO(d,1)$-invariant. 

The one-loop inner product~\eqref{E:twirledInnerProduct} implies that 
\beq
\label{E:resolution}
	\hat{\mathbb{1}} = \int \frac{[d\gamma]}{\text{diff}\times\text{Weyl}}|\gamma\rangle\!\rangle\langle\!\langle \gamma|\,, 
\eeq
at least when acting on asymptotic states $|\gamma'\rangle\!\rangle$ with $\gamma'$ close to the round metric. Here the measure implicitly carries a normalization constant $AS_0^{-\mathcal{D}_d/2}$ with $A^{-1} = |\mathcal{Z} \text{det}'(K)|$ with $|\text{det}'(K)|$ the ghost determinant with zero modes deleted (the zero modes correspond to the residual conformal isometry which we are treating by hand). The one-loop inner product proceeds mutatis mutandis to that presented in~\eqref{E:inversion}. Eq.~\eqref{E:resolution} implies that the contribution of one-sphere final states with $\gamma$ perturbatively close to the round sphere is
\beq
	\langle\!\langle \Psi|\Psi\rangle\!\rangle = \int \frac{[d\gamma]}{\text{diff}\times\text{Weyl}} |\Psi[\gamma]|^2\,.
\eeq

If we consider gravity coupled to a light scalar these results are only slightly modified. Asymptotic states can now be labeled as $|\mathfrak{h},\varphi\rangle\!\rangle$, and the one-loop inner product is
\beq
	\langle\!\langle \mathfrak{h,\varphi}|\mathfrak{h}',\varphi'\rangle\!\rangle = |\mathcal{Z}|S_0^{\mathcal{D}_d/2} \int_{SO(d,1)}\!\!\! d\alpha \,\langle \mathfrak{h},\varphi|\hat{U}(\alpha)|\mathfrak{h}',\varphi'\rangle
\eeq
with $|\mathcal{Z}|\langle \mathfrak{h},\varphi|\mathfrak{h}',\varphi'\rangle = |\mathcal{Z}|\delta[\mathfrak{h}-\mathfrak{h}']\delta[\varphi-\varphi']$ the joint ultralocal short-time amplitude obtained without integrating over the conformal twists. The contribution to the norm of a state $|\Psi\rangle\!\rangle$ from states with $\gamma$ perturbatively close to a round sphere is given by our main result~\eqref{E:main} with $A$ as above. 

\subsection{The no-boundary wavefunction and the absence of Polchinski's phase}
\label{S:noPhase}

We have argued for the expression~\eqref{E:main} as the contribution to the norm of a late-time state from configurations where $\gamma$ is close to being a round sphere. In particular the inner product of asymptotic states is non-negative. To verify that the norm of the no-boundary state is non-negative we need only check that it has no non-normalizable directions.  To proceed we require some basic features of the no-boundary wavefunction. 

In $d>3$ we can argue quite generally that the wavefunction has no wrong-sign directions. Extra input is required to ensure it has no bosonic zero modes. Consider a different problem, the Bunch-Davies wavefunction for a massive scalar field. We review this problem in the Appendix in some detail. When the scalar is stable, i.e.~has $M^2\geq 0$, it has no wrong-sign fluctuations on the sphere and the Bunch-Davies wavefunction on the equator, i.e.~at time $t=0$ it is a right-sign Gaussian distribution. (If the scalar is massless then the wavefunction has a zero mode, the constant mode of the scalar on the equator.) By unitarity these statements remain true at late time. 

For a tachyonic scalar now there are wrong-sign modes on the sphere indicating an instability to scalar condensation. These modes can be rotated to give a finite saddle-point contribution, leading to a phase in the sphere partition function. The instability on the sphere leads to certain wrong-sign directions in the Bunch-Davies wavefunction both at the equator and at late time. 

For gravity in de Donder gauge the late-time wavefunction only depends on the $\faktor{\text{TT}}{\text{LT}}$ modes. On the sphere these modes are all stable, and the technical problem of computing the late-time wavefunction at fixed angular momentum on the spatial sphere boils down to that of some stable scalars of some angular momentum and mass. To demonstrate that there are no zero modes we must show that the effective scalars are always massive and non-tachyonic.

It is interesting to explore how the wavefunction behaves in more detail.  We will discuss the cases of even $d$ and odd $d$ separately.

In even $d$ the wavefunction is invariant under diff$\times$Weyl. To evaluate it to leading non-trivial order in gravitational perturbation theory we require two ingredients. The first is the classical trajectory, the solution to the Einstein's equations that smoothly fills in the late-time boundary condition. Because we are only working to leading non-trivial order in small $G$, we need only consider $\gamma$ perturbatively close to the round metric $\gamma_0$, i.e.~$\gamma = \gamma_0 +\sqrt{G} \,\mathfrak{h}$. The second ingredient is the perturbative spectrum of fluctuations around the saddle above where $\gamma=\gamma_0$. The wavefunction is schematically\footnote{Here we have redefined the asymptotic states by $\gamma$-dependent phases so that the wavefunction has a finite late-time limit. This redefinition is the de Sitter version of the familiar holographic renormalization procedure in asymptotically AdS spacetimes, and can equivalently be understood by solving the Wheeler-de Witt equation perturbatively in the limit of large spatial universes~\cite{Chakraborty:2023yed}.}
\beq
\label{E:psiHH}
	\Psi_{\rm HH}[\mathfrak{h}] = \langle\!\langle\mathfrak{h}|\text{HH}\rangle \! \rangle \approx \Psi_{1\text{-loop}}  \,e^{ \frac{S_0}{2} - \sum_{\ell} \psi_{\ell} |\mathfrak{h}_{\ell}|^2 }\left( 1 + O(\sqrt{G})\right)\,,
\eeq
and some explanation is in order. We have decomposed the metric fluctuation $\mathfrak{h}$ into angular harmonics $\mathfrak{h}_{\ell}$ (where we are suppressing certain indices) for its transverse traceless directions on the late-time $\mathbb{S}^{d-1}$, the coefficients $\psi_{\ell}$ are computed from the classical trajectory mentioned above, $\Psi_{1\text{-loop}}$ is the one-loop determinant from fluctuations around the classical trajectory, and the first corrections of  $O(\sqrt{G})$ come from the cubic self-interactions of Einstein gravity. In $d=4$ the $\text{Re}(\psi_{\ell})$ are all known to be non-negative so that the round metric is a local maximum in the (unnormalized) probability distribution of metrics $|\Psi_{\rm HH}|^2$, behaving as a joint Gaussian distribution in the $\mathfrak{h}_{\ell}$'s. We gave an argument that this remains true for general $d$ above.

The simplest way to understand this fact is that late-time de Sitter wavefunctions behave like CFT partition functions in $d-1$ dimensions coupled to sources including a metric $\gamma_{ij}$. For example, the quadratic part of the wavefunction can be written as 
\begin{align}
\begin{split}
\label{E:integratedTT}
&\exp\!\left( - \sum_{\ell} \psi_{\ell} |\mathfrak{h}_{\ell}|^2\right) \\
& \qquad = \exp\left( - \frac{c_T}{2}\int d\Omega_1 d\Omega_2 \sqrt{\gamma_0(\Omega_1)} \sqrt{\gamma_0(\Omega_2)} \,\langle T^{ij}(\Omega_1)T^{k\ell}(\Omega_2)\rangle \mathfrak{h}_{ij}(\Omega_1) \mathfrak{h}_{k\ell}(\Omega_2)\right),
\end{split}
\end{align}
where $\Omega_1$ and $\Omega_2$ parameterize two points on a round $d-1$ sphere, $\langle T^{ij}(\Omega_1)T^{k\ell}(\Omega_2)\rangle$ is the standard, unit-normalized positive-semidefinite tensor structure parameterizing the two-point function of the stress tensor of a CFT on $\mathbb{S}^{d-1}$. The $\psi_{\ell}$'s are all controlled by this one quantity $c_T$. By our general argument above $\text{Re}(c_T)>0$ in any even $d$; this has been shown explicitly in $d=4$~\cite{Bobev:2016sap}, and using~\cite{Maldacena:2002vr} those authors have argued that this persists in even $d$. The damping of the wavefunction then follows. Note that we are not positing a dual CFT at future infinity. Rather the fluctuations of $\mathfrak{h}$ are governed by conformal symmetry on the sphere at late times in the same way as the integrated two-point function of a CFT stress tensor.

This presentation of the wavefunction makes another property of the wavefunction manifest, namely that all transverse traceless fluctuations are damped. The only modes that are not damped are the diff$\times$Weyl fluctuations of $\gamma$ thanks to the tracelessness and conservation of the effective stress tensor $T^{ij}$. So there are no bosonic zero modes of the distribution $|\Psi_{\rm HH}[\mathfrak{h}]|^2$ modulo diff$\times$Weyl, and combined with the damping of TT fluctuations, we see its norm is non-negative as claimed.

To one-loop the unnormalized probability distribution for $\mathfrak{h}$ is then the exponential of an integrated CFT two-point function,
\beq
	|\Psi_{\rm HH}[\mathfrak{h}]|^2 = |\Psi_{1\text{-loop}}|^2 e^{S_0 -2 \sum_{\ell}\text{Re}(\psi_{\ell})|\mathfrak{h}_{\ell}|^2}\left( 1 + O(\sqrt{G})\right)\,.
\eeq
The norm is the integral of this distribution over $\mathfrak{h}$ modulo diff$\times$Weyl. Thinking of the argument of the exponential as an action, the norm is an emergent conformal gravity path integral in an odd number $d-1$ of dimensions with a non-local effective action. The non-locality is unsurprising since there is no local and Weyl-invariant functional of $\gamma$ in odd dimension.

In odd $d$ the wavefunction behaves like an anomalous CFT partition function. In $d=3$ it is particularly simple, and the wavefunction is a constant after fixing $\gamma$ to be the round metric on the sphere. In general odd $d$ the anomalous variation of the on-shell action is real, meaning the effective anomaly coefficients are pure imaginary at tree level and so drop out of the unnormalized distribution $|\Psi_{\rm HH}[\mathfrak{h}]|^2$. As in $d=3$ it is plausible that the anomaly coefficients receive real corrections at loop level, in which case it must be that the measure at future infinity implied by the gravitational path integral cancels it in order for the norm to be consistent quantum mechanically. 

In extracting the late-time wavefunction we have allowed ourselves the freedom to rescale the asymptotic states $\langle\!\langle \gamma|$ by certain large phases corresponding to real counterterms built out of $\gamma$ and its derivatives on the late time slice. This procedure resembles holographic renormalization. In odd $d$ however there is now the feature that there are finite counterterms built from powers of the curvature. In the AdS/CFT correspondence these finite counterterms represent scheme-dependence and lead to scheme-dependent contact terms in stress tensor correlations. For example in $d=5$ there is a diff$\times$Weyl-invariant counterterm $\int d^4x \sqrt{\gamma} \,W(\gamma)^2$ with $W$ the Weyl tensor.

These terms are naturally generated in the no-boundary wavefunction with complex coefficients $C$. The real part can be removed by suitably redefining the external states, but the imaginary part survives. So the tree-level late-time wavefunction is
\begin{align}
\begin{split}
	\log \Psi_{\rm HH} \approx i S &= \frac{S_0}{2} - \frac{c_T}{2}\int d\Omega_1 d\Omega_2 \sqrt{\gamma_0(\Omega_1)}\sqrt{\gamma_0(\Omega_2)} \,\langle T^{ij}(\Omega_1)T^{kl}(\Omega_2)\rangle \mathfrak{h}_{ij}(\Omega_1)\mathfrak{h}_{kl}(\Omega_2) 
	\\
	& \qquad + i C \int d\Omega\sqrt{\gamma_0} \,\mathcal{W}(\mathfrak{h}) + O(\mathfrak{h}^3)\,,
\end{split}
\end{align}
where $c_T$ and $C$ are pure imaginary and the integrand in the second line is the quadratic approximation to a finite, real, Weyl-invariant Lagrangian built from $\gamma$, for example $\int d^4x \sqrt{\gamma}\,W(\gamma)^2$ in $d=5$. The distribution simplifies to
\beq
	\log |\Psi_{\rm HH}|^2\approx S_0 - 2\, \text{Im}(C) \int d\Omega \sqrt{\gamma_0}\,\mathcal{W}(\mathfrak{h}) + O(\mathfrak{h}^3)\,.
\eeq
This quantity is the ``action'' for the emergent conformal gravity path integral computing the late-time norm. By the argument at the beginning of this Subsection we expect this action to be maximized at $\mathfrak{h}=0$, and the only deformations that leave the action unchanged are those generated by diff$\times$Weyl transformations of $\mathfrak{h}$. 

\subsection{The norm vanishes at one loop}

In the last Subsection we argued that the late-time one-loop norm of the no-boundary state is non-negative. Moreover~\eqref{E:main} gives the one-loop approximation to the norm of a state $\Psi[\gamma,\varphi]$ peaked around $\gamma = \gamma_0$ as a gravitational path integral at future infinity. Using these results we now show that for the no-boundary state that approximation to the norm vanishes.

First consider pure gravity. Recall that the no-boundary wavefunction is approximated by a joint Gaussian distribution over fluctuations of the metric at future infinity around the round sphere. So the round sphere $\gamma = \gamma_0$ is a stable saddle point of the one-loop norm. To integrate over the fluctuations around it we must gauge fix the diff$\times$Weyl symmetry and this can be done via the Faddeev-Popov procedure by say directly fixing the non-transverse traceless fluctuations of the metric $\gamma$ to vanish. There are no bosonic zero modes of this gravity problem, which follows in even $d$ from the representation~\eqref{E:integratedTT} of the wavefunction as an integrated stress-tensor two-point function of a fictitious $d-1$-dimensional CFT on the sphere, and which we argued in odd $d$ above. However there are ghost zero modes corresponding to residual diff$\times$Weyl transformations that have not yet been fixed, the $SO(d,1)$ conformal isometries of the sphere. The volume of the gauge group then factorizes into a product
\beq
	V_{\text{diff}\times\text{Weyl}} = V_{\rm FP} \times V_{\rm residual}\,,
\eeq
with $V_{\rm FP}$ the part that is fixed through the nonzero modes of the Faddeev-Popov ghosts and $V_{\rm residual}$ the volume of the residual $SO(d,1)$ symmetry. Dividing by gauge transformations to one-loop order means that we only integrate over the bosonic modes, the nonzero ghost modes, and divide by the volume of the residual symmetry. This gives the result we claimed in the Introduction,
\beq
	\langle\!\langle \text{HH}|\text{HH}\rangle\!\rangle = e^{S_0} \frac{\widetilde{Z}S_0^{-\mathcal{D}_d/2}}{\text{vol}(SO(d,1))} + (\text{higher loops, other geometries})\,.
\eeq
Here $\widetilde{Z}$ is a complicated determinant that includes contributions from (1) the integral over the TT modes of $\gamma$ in the gravitational path integral at future infinity, (2) the inverse normalization of the inner product~\eqref{E:twirledInnerProduct}, and (3) the one-loop determinants that contribute to the wavefunction normalization $\Psi_{1\text{-loop}}$. 

Adding free matter does not change the qualitative form of this result. To one-loop order the late-time wavefunction factorizes into a product
\beq
	\langle \!\langle \mathfrak{h}, \varphi|\text{HH}\rangle\!\rangle = \Psi_{\rm HH}[\mathfrak{h}] \Psi_{\rm BD}[\varphi]\,,
\eeq
and it is still $SO(d,1)$ invariant. So the argument above still goes through with
\beq
	\langle \!\langle \text{HH}|\text{HH}\rangle\!\rangle =e^{S_0}\frac{\widetilde{Z}S_0^{-\mathcal{D}_d/2}}{\text{vol}(SO(d,1))} \langle \text{BD}|\text{BD}\rangle + (\text{higher loops, other geometries})\,.
\eeq

\subsection{Overlaps in $d>3$ beyond one loop}
\label{subsec:higherloops}

So far we have discussed the inner product on asymptotic states at one loop, coming from the measure $\frac{[d\gamma]}{\text{diff} \times \text{Weyl}}$.  Crucially, the conformal Killing vectors, which furnish a representation of $SO(d,1)$, are residual gauge symmetries of e.g.~the transverse traceless gauge fixing condition on $\gamma$ at one loop.  Then as explained above, the mechanism for the vanishing of the norm of the Hartle-Hawking state at one loop is that it is invariant under $SO(d,1)$ but does not possess bosonic zero modes; as such, the fixing of the residual $SO(d,1)$ gauge symmetry leads to the norm being proportional to $1/\text{vol}(SO(d,1))$.  In this Subsection we explain the residual gauge symmetry in a slightly different language, and extend the result to all loops.

Let $\gamma_{ij} = \gamma_{0\,ij} + \sqrt{G}\,\mathfrak{h}_{ij}$ where $\gamma_{0\,ij}$ is the round metric on $\mathbb{S}^{d-1}$ as before. To fix $\gamma_{ij}$ modulo diff$\times$Weyl, we impose the gauge-fixing conditions (analogous to those in~\cite{Chakraborty:2023los})
\begin{align}
F_i[\gamma] \equiv \nabla_{\gamma_0}^j \mathfrak{h}_{ij}  = 0\,,\quad H[\gamma] \equiv \gamma_0^{ij}  \mathfrak{h}_{ij} = 0\,,
\end{align}
where $\nabla_{\gamma_0}$ denotes the covariant derivative constructed from $\gamma_0$.
Here $\nabla_{\gamma_0}^j = \gamma_0^{ij}\nabla_{\gamma_0\,j}$. Then the diffeomorphism and Weyl variations of $F_i$ and $H$ give
\begin{align}
\delta_{\xi, \sigma} F_i &= \nabla_{\gamma_0}^j( \nabla_{\gamma\,i} \xi_j + \nabla_{\gamma\,j} \xi_i + 2 \sigma \gamma_{ij} ) \\
&=  \nabla_{\gamma_0}^j\left(\nabla_{\gamma_0\,i} \xi_j + \nabla_{\gamma_0\,j} \xi_i  + 2 \gamma_{0\,ij}  \sigma\right) + \sqrt{G} \,\nabla_{\gamma_0}^j\left(- \gamma^{k\ell} (\nabla_{\gamma_0\,i} \mathfrak{h}_{j\ell} + \nabla_{\gamma_0\,j} \mathfrak{h}_{i\ell} - \nabla_{\gamma_0\,\ell} \mathfrak{h}_{ij}) \xi_k + 2 \mathfrak{h}_{ij} \sigma\right) \nonumber
\end{align}
and
\begin{align}
\begin{split}
\delta_{\xi, \sigma} H &= \gamma_0^{ij}(  \nabla_{\gamma\,i} \xi_j + \nabla_{\gamma\,j} \xi_i + 2 \sigma \gamma_{ij} ) \\
&= \gamma_0^{ij}\left( \nabla_{\gamma\,i} \xi_j + \nabla_{\gamma\,j} \xi_i) + 2  (d-1) \sigma\right) .
\end{split}
\end{align}
Observe that to $O(1)$ in $\sqrt{G}$, we can simultaneously solve $\delta_{\xi, \sigma} F_i = 0$ and $\delta_{\xi, \sigma} H = 0$ by letting $\xi_i = \mathcal{K}_i$ and $\sigma = - \frac{1}{d-1} (\nabla_{\gamma_0} \cdot \mathcal{K})$, where $\mathcal{K}_i$ is a conformal Killing vector on $\mathbb{S}^{d-1}$.  Thus we see in another way that the conformal Killing vectors furnish residual gauge symmetries in the quadratic approximation to the norm, as we previously discussed in Section~\ref{subsec:overlaps1}.

We presently extend that analysis to higher loops by finding perturbative corrections to $\xi_i = \mathcal{K}_i$ so that $\xi_i$ generates a residual symmetry to all orders in $\sqrt{G}$.  First we note that $\delta_{\xi, \sigma} H = 0$ gives
\begin{align}
\sigma = - \frac{1}{d-1}(\nabla_{\gamma_0} \cdot \xi)
\end{align}
and so plugging this into $\delta_{\xi, \sigma} F_i = 0$ we find
\begin{align}
\begin{split}
\label{E:xieq1}
&\nabla_{\gamma_0}^j\left(\nabla_{\gamma_0\,i} \xi_j + \nabla_{\gamma_0\,j} \xi_i  - \frac{2}{d-1} \gamma_{0\,ij}  (\nabla_{\gamma_0} \cdot \xi)\right)  \\
& \qquad \qquad = - \sqrt{G}\, \nabla_{\gamma_0}^j\left(-  (\nabla_{\gamma_0\,i} \mathfrak{h}_{j\ell} + \nabla_{\gamma_0\,j} \mathfrak{h}_{i\ell} - \nabla_{\gamma_0\,\ell} \mathfrak{h}_{ij}) \gamma^{k\ell} \xi_k - \frac{2}{d-1}\, \mathfrak{h}_{ij}  (\nabla_{\gamma_0} \cdot \xi)\right).
\end{split}
\end{align}
Suppose we decompose $\xi_i = \sum_{p = 0}^\infty G^{p/2}\,\xi_i^{(p)}$ where $\xi_i^{(p)}$ is $p$th order in $\mathfrak{h}_{ij}$, and similarly $\gamma^{ij} = \sum_{p = 0}^\infty G^{p/2}\,[\gamma^{k\ell}]^{(p)}$ where $[\gamma^{k\ell}]^{(p)}$ denotes the contribution to $\gamma^{k\ell}$ which is $p$th order in $\mathfrak{h}_{ij}$.  Then we find that~\eqref{E:xieq1} becomes
\begin{align}
\label{E:xieq2}
&\nabla_{\gamma_0}^j\left(\nabla_{\gamma_0\,i} \xi_j^{(p)} + \nabla_{\gamma_0\,j} \xi_i^{(p)}  - \frac{2}{d-1} \gamma_{0\,ij}  (\nabla_{\gamma_0} \cdot \xi^{(p)})\right)  \\
& \qquad \quad = -  \nabla_{\gamma_0}^j\left(-  (\nabla_{\gamma_0\,i} \mathfrak{h}_{j\ell} + \nabla_{\gamma_0\,j} \mathfrak{h}_{i\ell} - \nabla_{\gamma_0\,\ell} \mathfrak{h}_{ij}) \sum_{q=0}^{p-1}[\gamma^{k\ell}]^{(p-1-q)} \xi_k^{(q)} - \frac{2}{d-1}\, \mathfrak{h}_{ij}  (\nabla_{\gamma_0} \cdot \xi^{(p-1)})\right). \nonumber
\end{align}
for $p = 0,1,2,...$.  We now show that we can systematically solve for $h$-corrections to the conformal Killing vectors.

Writing the left-hand side of~\eqref{E:xieq2} as $\textsf{D}_i[\xi_j^{(p)}]$, we note that $\textsf{D}_i$ is an elliptic differential operator on a compact manifold.  Its only zero eigenvectorfields are the conformal Killing vectors.  The Fredholm alternative theorem ensures a unique solution for $\xi^{(p)}$ if the right-hand side of~\eqref{E:xieq2} is orthogonal to the kernel of $\textsf{D}_i$.  Indeed, integrating the right-hand side of~\eqref{E:xieq2} against a conformal Killing vector $\mathcal{K}^i$ and integrating by parts in the $\nabla_{\gamma_0}^j$, we find
\begin{align}
\int_{\mathbb{S}^{d-1}} \!\! d^{d-1}\Omega \sqrt{\gamma_0} \,\nabla_{\gamma_0}^j \mathcal{K}^i \!\!\left(\!\!-  (\nabla_{\gamma_0\,i} \mathfrak{h}_{j\ell} + \nabla_{\gamma_0\,j} \mathfrak{h}_{i\ell} - \nabla_{\gamma_0\,\ell} \mathfrak{h}_{ij}) \!\sum_{q=0}^{p-1}[\gamma^{k\ell}]^{(p-1-q)} \xi_k^{(q)} \!-\! \frac{2}{d-1}\, \mathfrak{h}_{ij}  (\nabla_{\gamma_0} \cdot \xi^{(p-1)})\!\!\right).
\end{align}
But the above vanishes by using the conformal Killing equation, in tandem with the gauge fixing conditions for $\mathfrak{h}_{ij}$.  As such, the Fredholm alternative theorem applies, and there is a unique solution for each $\xi_i^{(p)}$.

In summary, we have shown that to all orders in metric fluctuations, the transverse traceless gauge fixing condition on $\gamma$ has residual gauge transformations corresponding to deformations of the conformal Killing vectors on $\mathbb{S}^{d-1}$. This establishes that a crucial ingredient in the vanishing of the no-boundary inner product at one loop persists to all loops.  Specifically, if the loop-corrected Hartle-Hawking state were to be invariant under the deformed symmetry and also have no bosonic zero modes, then upon computing its norm the fixing of the residual symmetry would lead to division by an infinite volume.

\subsection{Some comments}

\begin{figure}[t!]
\begin{center}
\includegraphics[scale=.45]{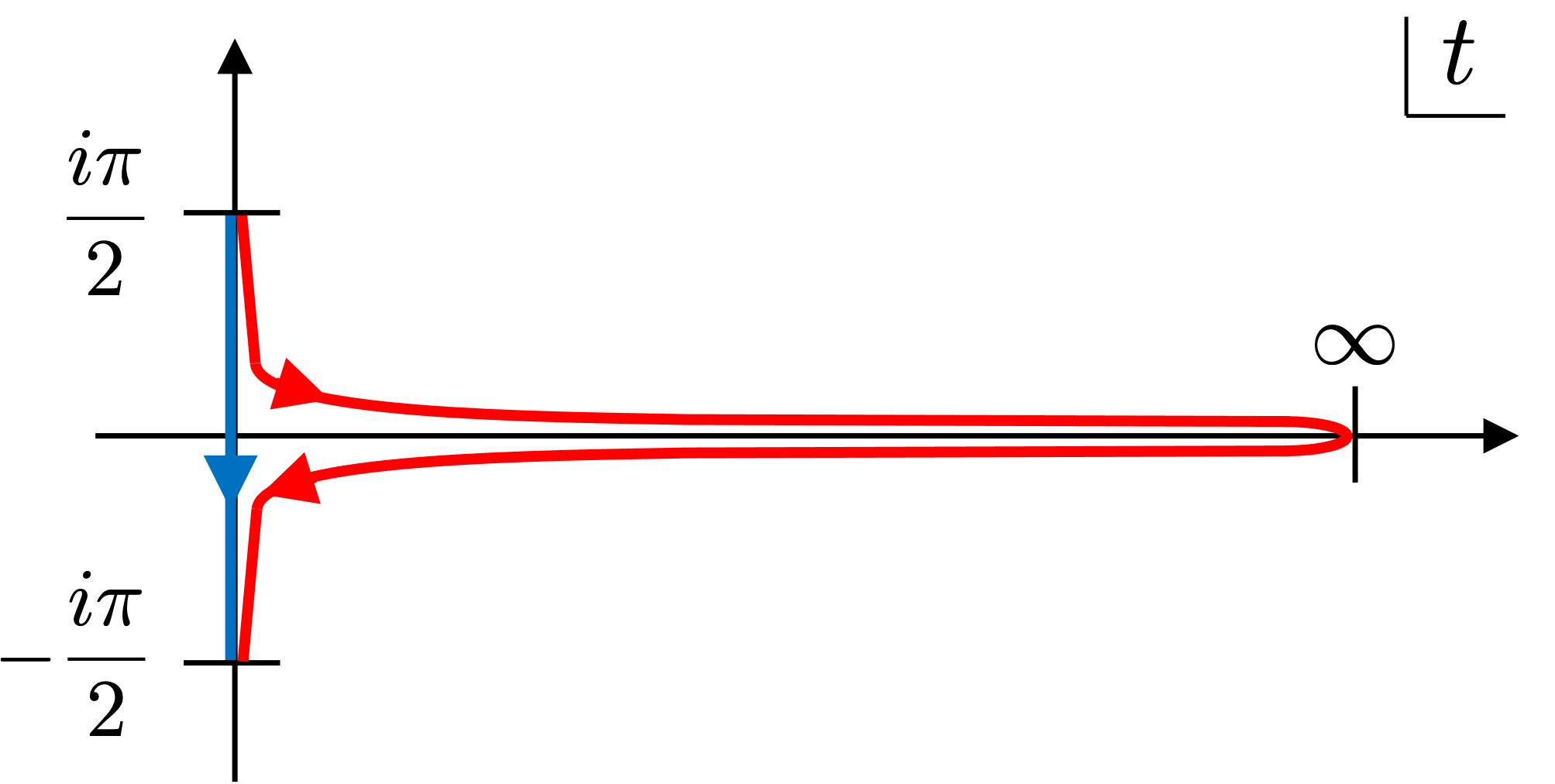}
\end{center}
\caption{The complex time contour for the inner product of the no-boundary state with itself at late time (red), and the one (blue) for the sphere amplitude. \label{Fig:contour1}}
\end{figure}

As we have seen the norm of the no-boundary state can be recast as a gravitational path integral with a complex time contour that has a turnaround point at late Lorentzian time. In field theory such a path integral can be smoothly deformed from one on that contour to one on the Euclidean sphere. However for gravity we have argued that this is not the case, with the former being non-negative and the latter having a dimension-dependent phase. See Fig.~\ref{Fig:contour1}.

We would like to gain physical understanding for what distinguishes these two computations. We have found three qualitative differences worth pointing out.

The first is that in our Lorentzian result we glue together the wavefunction and its complex conjugate, i.e.~stitching together a bra and a ket, whereas the Euclidean path integral on the sphere is more naturally a pairing between the no-boundary state and itself, rather than an overlap between bra and ket. 

The second is that both on the sphere and at future infinity there is a residual gauge symmetry that we divide by. For the sphere this is its $SO(d+1)$ isometry, whereas for the late-time norm it is the conformal isometry $SO(d,1)$ of the late-time sphere. We note that if we analytically continued the $SO(d+1)$ to act on the entire complex time contour that computes the norm, it would act as a complex diffeomorphism at late time. Forcing it to act in a real way at the turn-around point amounts to analytically continuing it to $SO(d,1)$, the residual symmetry of the Lorentzian norm.

Lastly, we note that there is an analogue of Polchinski's phase for tachyonic scalar theories, which in that case encodes physics. To wit, the ensuing Bunch-Davies wavefunction is a wrong-sign Gaussian in directions corresponding to the wrong-sign modes of the tachyonic scalar on the sphere. This non-normalizability is visible both for the wavefunction at $t=0$, i.e.~the hemisphere path integral, and by unitarity at late time. One may evaluate the ``norm'' of the Bunch-Davies state for the tachyon by rotating the contour of integration for those modes at the cost of generating a phase which for a choice of contour rotation reproduces the phase of the sphere partition function.

Na\"{i}vely this result, which we recapitulate in the Appendix, is relevant in gravity as the rotated conformal mode of the metric behaves to one-loop order like a tachyonic scalar with mass-squared $M^2 = -2(d-1)$. For that mass-squared the late-time Bunch-Davies wavefunction is non-normalizable in the $\ell=1$ directions, meaning the $\ell=1$ fluctuations on the sphere at late time. However and crucially, for the gravity problem the wavefunction does not depend on the late-time limit of the conformal mode. Indeed, the late-time probability distribution $|\Psi_{\rm HH}|^2$ is a right-sign joint Gaussian distribution in physical metric fluctuations as discussed in and around Eq.~\eqref{E:psiHH}. This fact is vividly clear for the case of pure 3d gravity where there are no physical fluctuations of the metric $\gamma$ period.

This gives an intuitive argument for the non-negativity of the one-loop norm of the no-boundary state. In a sentence, unlike for a tachyonic scalar, there are no non-normalizable directions in the late-time wavefunction of the no-boundary state of de Sitter gravity.

\section{Adding insertions}
\label{S:matter}

So far we have studied the one-loop approximation to the norm of the no-boundary state in pure Einstein gravity and in models of gravity coupled to matter. The one-loop contribution is proportional to $1/\text{vol}(SO(d,1))$ stemming from a residual gauge symmetry of the norm, namely the conformal isometry of the sphere, and so vanishes. However, in analogy with string amplitudes, we can stabilize the overlap to a nonzero value with appropriate insertions. We explore some examples below, focusing on the endpoints of worldlines with clocks as well as non-Gaussian corrections to the no-boundary wavefunction. We note in passing that `handles' or other non-trivial topological features at future infinity also yield finite results. In particular, contributions to the no-boundary state where future infinity is some other space in which the conformal isometry $SO(d,1)$ is broken to a compact subgroup (e.g.~$\mathbb{S}^1 \times \mathbb{S}^{d-2}$, $\mathbb{T}^{d-1}$, etc.) give finite contributions to the norm.

\subsection{Introducing a worldline with a clock}

Consider adding a worldline with a clock to de Sitter, which can be regarded as a toy model of an observer.  We utilize the observer model of~\cite{Witten:2023xze} described by the action
\begin{align}
\label{E:NG1}
S_{\text{obs}}[X^\mu, q,p] = \int d\tau\left(p \frac{dq}{d\tau} - \sqrt{-g_{\tau\tau}} (m+q)\right)\,,
\end{align}
where we take $q \geq 0$ and $p\in \mathbb{R}$.  Above, $\tau$ is the proper time of the worldline and $g_{\tau \tau}$ is the induced metric. Moreover $q$ serves as a `clock' degree of freedom and the observer's Hamiltonian $H_{\text{obs}} = m + q$ indicates a lower bound on its energy $E\geq m$. The variable $p$ is the `time' variable and its noncompactness leads to a continuous, uniform spectrum for $q$. On account of this continuous spectrum the observer carries an infinite entropy.

 It is useful to work with the Polyakov form of~\eqref{E:NG1},
\begin{align}
\label{E:Polyakov1}
S_{\text{obs}}[X^\mu, q,p,e]  = \int d\tau\left(p \frac{dq}{d\tau} + \frac{1}{2e}(m+q) g_{\mu \nu} \dot{X}^\mu \dot{X}^\nu - \frac{e}{2}(m+q) \right),
\end{align}
where we have introduced an einbein $e(\tau)$. Let us first compute the wavefunction of the no-boundary state including the worldline. To one-loop order the wavefunction factorizes into the product of the gravitational part $\Psi_{\rm HH}[\mathfrak{h}]$ and the wavefunction for the observer $\Psi_{\rm obs}$  in the absence of gravity. It is impossible for the worldline to end at only a single point on the boundary. Instead it must connect two points at late time $t=T$. Gauge-fixing $e = 1$ we can write the wavefunction of the observer as
\begin{align}
\label{E:Psiobs1}
\Psi_{\text{obs}}(\Omega_1, q_1, \Omega_2, q_2) = \int [dX^\mu][dq] [dp][db][dc]\,e^{i S_{\rm obs}}\,,
\end{align}
where $b,c$ form a bc ghost system with ghost numbers $(-1,+1)$. The integral is taken over trajectories on the Hartle-Hawking geometry that connect a point on the late-time slice $(T,\Omega_1)$ at which $q = q_1$, to another point $(T,\Omega_2)$ at which $q=q_2$, in the limit $T\gg 1$.  

\begin{figure}
\begin{center}
\includegraphics[scale=.6]{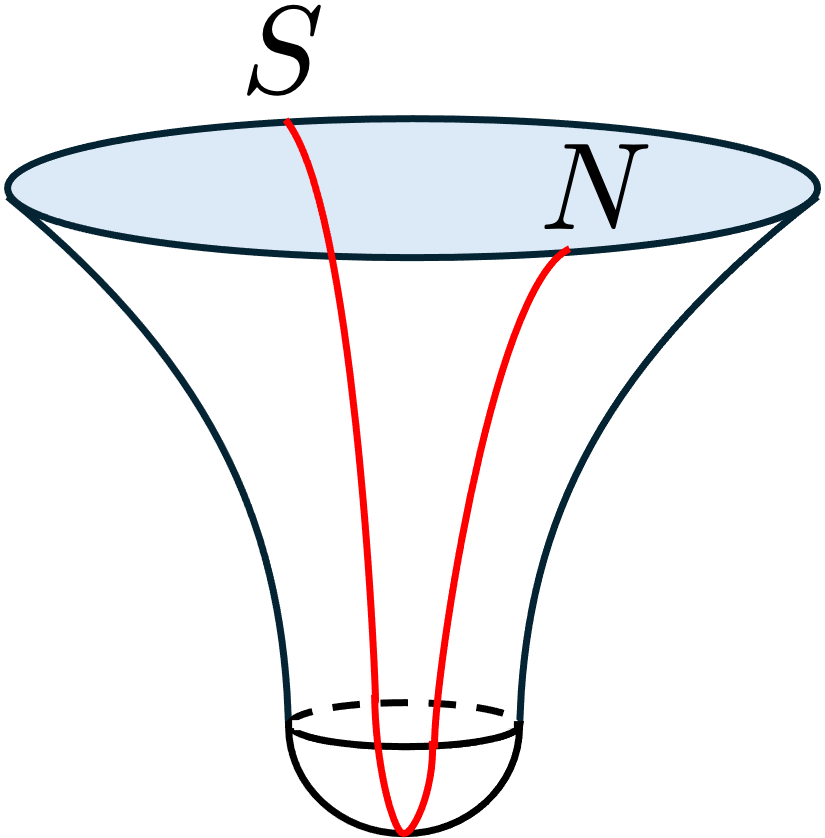}
\end{center}
\caption{The geodesic connects the south pole $S$ of the sphere in the far future to the north pole $N$ through a trajectory that passes through the Euclidean cap. \label{fig:HHworldline}}
\end{figure}

We can use a subset of the conformal isometries to fix $\Omega_1$ to be the south pole of the sphere and $\Omega_2$ to be the north pole.  The geodesic connecting those two points passes through the pole of the Euclidean cap as in Fig.~\ref{fig:HHworldline} and breaks the $SO(d,1)$ isometry to $SO(1,1) \times SO(d-1)$. Then
\begin{align}
\label{E:psiObserver}
\Psi_{\text{obs}}(q_1, q_2) = e^{-\pi (m + q_1) } \,\delta(q_1-q_2)\,\mathcal{J}_{\text{obs}}^{1\text{-loop}} \,,
\end{align}
where $\mathcal{J}_{\text{obs}}^{1\text{-loop}}$ is a one-loop determinant over nonzero modes of the observer degrees of freedom. Because we can use the $SO(d,1)$ isometry to move the two insertions to any two non-coincident points on the late-time sphere, this is in fact the wavefunction $\Psi_{\text{obs}}(\Omega_1,q_1, \Omega_2, q_2)$ as a function of the energies and the final positions $\Omega_1$ and $\Omega_2$.

With the observer, the completeness relation now includes integrals over the positions of the endpoints
\begin{align}
\hat{\mathbb{1}} \!\propto\! \int \frac{[d\gamma]}{\text{diff}\times\text{Weyl}} \!\left( \int d^{d-1}\Omega_1 dq_1\sqrt{\gamma(\Omega_1)}\right)\!\!\left(  \int d^{d-1}\Omega_2dq_2 \sqrt{\gamma(\Omega_2)} \right)\!|\gamma;\Omega_1,q_1,\Omega_2,q_2\rangle\!\rangle\langle \!\langle \gamma;\Omega_1,q_1,\Omega_2,q_2|\,.
\end{align}
In order for this measure to be viable it must be the case that the observer wavefunction~\eqref{E:psiObserver} is dressed suitably, $|\Psi_{\rm obs}|^2 \to |\widetilde{\Psi}_{\rm obs}|^2$, so that $\int d^{d-1}\Omega_1 \sqrt{\gamma(\Omega_1)}\int d^{d-1}\Omega_2 \sqrt{\gamma(\Omega_2)} |\widetilde{\Psi}_{\rm obs}|^2$ is Weyl-invariant. This is a similar problem to understanding how the field operators appearing in cosmological correlators can be upgraded to gauge-invariant observables when integrating over data mod diff$\times$Weyl, one whose outcome we do not presently understand but which we assume can be solved.

Assuming such a dressing exists the norm of the state with the worldline then behaves like the sphere approximation to the two-point function in string theory, with 
\beq
\label{E:innerproductworldline1}
\langle \!\langle\text{HH}|\text{HH}\rangle\!\rangle = e^{S_0} \frac{\widetilde{Z}S_0^{-\mathcal{D}_d/2}}{\text{vol}(SO(1,1))\times\text{vol}(SO(d-1))} \times \left( e^{- 2\pi m} m^{-(d-1)}\,Z_{\text{obs}}\,\delta(0) \right)\,. 
\eeq
The first part is the gravitational contribution while the term in parentheses comes from the observers. We are neglecting higher loops and other geometries at future infinity. The first term therein $e^{-2\pi m}= e^{- \beta_{\text{dS}} m}$ agrees with the analysis of~\cite{Chandrasekaran:2022cip, Witten:2023xze}, and the second $Z_{\text{obs}}$ is a positive one-loop correction.  The $\delta(0) = \text{vol}(\mathbb{R})$ comes from the energy (or $q$) zero mode, namely the volume of $p$, and can be thought of as an infinite observer entropy. This expectation value is a ratio of divergent quantities, $\propto \frac{\delta(0)}{\text{vol}(SO(1,1))}$, but since both of these quantities are $\text{vol}(\mathbb{R})$ we expect they cancel up to a finite $m$-independent ratio; we absorb this ratio as well as $Z_{\rm obs}$ into $\widetilde{Z}$ giving
\beq
\label{E:HHwithobs1}
\langle \!\langle\text{HH}|\text{HH}\rangle\!\rangle  = e^{S_0 - 2 \pi m} \frac{\widetilde{Z}'S_0^{-\mathcal{D}_d/2}m^{-(d-1)}}{\text{vol}(SO(d-1))}
\eeq
at one loop. This leads to a finite inner product~\eqref{E:innerproductworldline1}, and so the presence of two observers with clocks ameliorates the division by $\text{vol}(SO(d,1))$. 

At small $G$ and large $m =O(1/G)$ this result is $\sim e^{S_0 - 2\pi m}S_0^{-(\mathcal{D}_{d-2}+1)/2}$ whose logarithm is the classical approximation to the de Sitter horizon entropy in the presence of the worldline.  Since this norm is non-negative and reproduces the classical entropy we hypothesize that it defines the entropy quantum mechanically.

As remarked in the Introduction, the finiteness of~\eqref{E:HHwithobs1} relies on a model of an observer with infinite entropy, as per~\cite{Witten:2023xze}, to soak up the $\text{vol}(SO(1,1))$.  Conversely, models of observers with finite entropy (see e.g.~\cite{Maldacena:2024spf}) would lead to a vanishing norm.  While we do not yet know the moral of this finding, we speculate that it is telling us that an observer that survives to the infinite future must have infinite temporal resolution and therefore entropy.

\subsection{Matter and inflatons}

As we previously discussed, the no-boundary state including matter in the Bunch-Davies state has zero norm at one loop. Suppose instead of computing the norm we consider the one-loop approximation to unnormalized correlation functions of appropriate gauge-invariant operators, along the lines of the proposal of~\cite{Chakraborty:2023los} for fully gauge-invariant cosmological correlators. Certainly with three such insertions the residual $SO(d,1)$ symmetry is effectively broken down to a compact subgroup, rendering the correlator finite. In analogy with string amplitudes we expect two-point functions to also be finite. That being said, at present it is not clear what the fully gauge-invariant versions of field operators are, and how they align with standard (not fully gauge-invariant) methods for computing cosmological correlators.

Of physical interest is the no-boundary state including a slowly rolling inflaton in $d = 4$.  For simplicity, we consider a single scalar field $\phi$ with potential $V(\phi)$ in the slow-roll approximation $\varepsilon = \frac{G}{2} \left(\frac{V'}{V}\right)^2 \ll 1$ and $|\eta| = G \frac{|V''|}{V} \ll 1$.  Following e.g.~\cite{Chen:2024rpx}, it is convenient to consider the probe regime for the inflaton on top of an asymptotically de Sitter background, where we take $V \sim O(1/G)$, $V' \sim O(1)$, and $V'' \sim o(1)$ for $G$ small.  Here $\Lambda = 8 \pi G V$ on distance scales much larger than the de Sitter length.

In this approximation the wavefunction factorizes at one loop into the product of a gravitational wavefunction and that of the inflaton. Both are $SO(4,1)$ invariant.  (See~\cite{Maldacena:2024uhs, Chen:2024rpx} for a recent discussion.)  The $SO(4,1)$ invariance of the slow-roll wavefunction is due to the fact that the modes with non-zero angular momenta on the spatial sphere have a quadratic kernel which is effectively that of a massless scalar, and the remaining zero mode, $\varphi_{\ell,\textbf{m}} = \varphi_{0,\textbf{0}}$, appears linearly in the logarithm of the wavefunction but is inert under $SO(4,1)$.  In conventions where $V' > 0$ we have $\Psi[\gamma, \varphi] \sim e^{- CV' \varphi_{0,\textbf{0}}}$ for $C>0$ and so there is a pressure for $\varphi_{0,\textbf{0}}$ to be lower in the potential leading to fewer $e$-folds of inflation.  The $SO(4,1)$ invariance of the wavefunction implies that $\langle \! \langle \text{HH} | \text{HH} \rangle \! \rangle = 0$ at one loop.  As such, the leading contribution to the normalized cosmological correlators comes from higher loops.

Although from the tree level wavefunction $\varphi_{0,\textbf{0}}$ appears to be pressured to be lower in its potential, it would be interesting to compute suitably gauge-invariant correlation functions which probe this effect so as to determine if this is true quantum mechanically.  The basic issue is that $\varphi_{0,\textbf{0}}$ is not itself gauge-invariant, so some other probe is required.

We remark that inflationary cosmologists often consider the Bunch-Davies state in the inflating patch, i.e.~flat slicing, of de Sitter space, described by $-dt^2 + e^{2t} d\textbf{x}^2$ for $t \in \mathbb{R}$ and $\textbf{x} \in \mathbb{R}^3$.  This is a geodesically incomplete patch of global de Sitter for which we must supply boundary conditions on the past horizon. Since the Hartle-Hawking geometry is geodesically complete and has compact spatial slices, one might regard it as more theoretically natural. In any case, in flat slicing, we can also consider a boundary metric $\gamma$ on the $\mathbb{R}^3$ at future infinity; then when computing norms including fluctuations of $\gamma$, it is reasonable to expect that we integrate over the data at future infinity modulo diff$\times$Weyl, but perhaps we should only gauge those transformations that go to the identity as $|\textbf{x}| \to \infty$.  (This is akin to how in quantum gravity in flat space we do not e.g.~gauge rotations.)  If that is the correct procedure then in this context there will be no division by $SO(4,1)$, nor is the classical approximation to the norm equal to $e^{S_0}$.

\subsection{Wavefunctions beyond one loop}

What is the leading nonzero contribution to the norm of the no-boundary state? In this manuscript we raise the prospect that the sphere contribution to the norm vanishes. Whether it actually vanishes is of course an important question.  Here we offer some additional observations and informed speculations that we hope will be useful to answer it in the future. 

As a warm-up, consider the Bunch-Davies wavefunction for an interacting quantum field theory, say $\phi^4$ theory. If the interactions are perturbative, characterized by a small parameter $\lambda$, then the wavefunction has the general form
\beq
	\Psi_{\rm BD}[\varphi] = \Psi_{1\text{-loop}} \, e^{- \sum_{\ell} \Upsilon_{\ell}|\varphi_{\ell}|^2} \left( 1 + O(\lambda)\right),
\eeq
where $\ell$ indicates the angular momentum on the sphere at future infinity, $\text{Re}(\Upsilon_{\ell})\geq 0$, and the corrections are cubic or higher in powers of $\varphi_{\ell}$.\footnote{Implicitly we define the theory by tuning a linear term so that $\varphi = 0$ remains the maximum value of $|\Psi|^2$, and absorb constant and quadratic corrections into the prefactor $\Psi_{1\text{-loop}}$ and the coefficients $\Upsilon_{\ell}$ respectively.} Conformal invariance of the wavefunction implies that they behave like integrated multi-point functions. Na\"ively these perturbative corrections to the wavefunction function as integrated vertex operators, so that in the norm, where in the non-gravitational limit we integrate over $\varphi$ mod $SO(d,1)$, we can fix the non-compact part of the $SO(d,1)$ symmetry by fixing three of the scalar insertions. By this argument we would expect that $\int \frac{[d\varphi]}{SO(d,1)} |\Psi_{\rm BD}|^2$ is nonzero. However this cannot be true, since on general grounds the ordinary norm of the Bunch-Davies state $\int [d\varphi]|\Psi_{\rm BD}|^2$ is the renormalized sphere partition function of the field theory which remains finite in the presence of renormalizable interactions.

This being said we suspect that the moral of this argument is that perturbative corrections to the wavefunction exponentiate. Thinking of the wavefunction as the exponential of an action, these corrections simply renormalize the `coupling constants.'  Then the corrected  $\int [d\varphi]|\Psi_{\rm BD}|^2$ would be the renormalized sphere partition function of the field theory.  This is analogous to the treatment of anomalous scaling dimensions in AdS/CFT~\cite{Freedman:1998bj}, and also is known to manifest in de Sitter in the context of the IR divergence problem~\cite{Gorbenko:2019rza}.

For example, with matter coupled to gravity, there are non-Gaussian corrections to the Bunch-Davies wavefunction when the metric at future infinity is the round one plus a small fluctuation $\gamma = \gamma_0+\sqrt{G} \,\mathfrak{h}$, which on general grounds implies
\begin{align}
	\mathcal{I}[\mathfrak{h}] &= \int [d\varphi] |\Psi_{\rm BD}[\varphi;\mathfrak{h}]|^2 
	\\
	\nonumber
	&= Z_{\rm matter}\!\left( 1 + \frac{G}{2}\int d\Omega_1 d\Omega_2 \sqrt{\gamma_0(\Omega_1)}\sqrt{\gamma_0(\Omega_2)}\,\langle T_{\varphi}^{ij}(\Omega_1)T_{\varphi}^{k\ell}(\Omega_2)\rangle_{\rm BD} \mathfrak{h}_{ij}(\Omega_1) \mathfrak{h}_{k\ell}(\Omega_2) + O(G^{3/2})\right).
\end{align}
This correction is a conformally invariant, integrated two-point function of the matter stress tensor at future infinity in the (normalized) Bunch-Davies state.\footnote{In odd $d$ this quadratic correction is actually local by the argument in Subsection~\ref{S:noPhase}.} A series of corrections of the above form should exponentiate to renormalize the effective Newton's constant in the Hartle-Hawking wavefunction, i.e.~the coefficient of the $O(\mathfrak{h})^2$ term in its logarithm.  Similarly, there are other corrections to the wavefunction coming from the pure gravity sector going like integrated three- and higher-point functions of metric fluctuations $\mathfrak{h}$. 

In Subsection~\ref{subsec:higherloops} we found that there is a residual gauge symmetry of the computation of the norm to all orders in fluctuations around the round metric on the sphere. We can think of it as a deformed version of the $SO(d,1)$ conformal isometry. It is natural to speculate that the no-boundary wavefunction is invariant under it.  While this seems plausible, it should be checked directly in some examples or confirmed by an abstract argument.

Here we have given some general, qualitative comments. We stress that detailed computation is required to settle the leading nonzero contribution to the norm. That result is physically important as it determines the normalization of cosmological correlators in the no-boundary state, or more dramatically, whether such correlators can be normalized.

\section{Discussion}
\label{S:discussion}

In this manuscript we have shown that the one-loop approximation to the norm of the no-boundary vanishes, even when coupled to matter in a conformally invariant state.  This includes the case of coupling to a slow-roll inflaton, and so our findings may be consequential for real-world settings.  We have also provided evidence that the sphere contribution to the no-boundary state may have vanishing norm, although additional computation and arguments will be required to reach a final verdict.  Even so, unnormalized cosmological correlators can be finite (e.g.~a three point function of vertex operators)~\cite{Chakraborty:2023yed, Chakraborty:2023los}, although the right set of gauge-invariant observables is presently unclear. But if the norm of the no-boundary state indeed vanishes to all loops, then the status of unnormalized cosmological correlators is puzzling.  We remark that while it is reasonable to expect that the (dominant contribution to the) wavefunction of the universe has asymptotically de Sitter ‘cosmological’ symmetries, our results run counter to this intuition since apparently the Hartle-Hawking contribution to the no-boundary wavefunction is null to at least one loop. 

Below we comment on some other features and questions brought forth by our results.
\\ \\
\textbf{Other topologies.} Although the Hartle–Hawking geometry, with its spherical spatial slices, would seem to give the dominant contribution to $|\Psi_{\rm HH}|^2$, there are contributions from other topologies at future infinity, such as $\mathbb{T}^{d-1}$, $\mathbb{S}^1 \times \mathbb{S}^{d-2}$, etc., which should be summed over in the path integral (see e.g.~\cite{Castro:2011xb, Castro:2012gc, Cotler:2024xzz, Godet:2025bju, Anninos:2025ltd}). These other topologies would become much more important if the sphere contribution to the norm of the no-boundary state vanishes. Similar techniques as the ones introduced in this paper should apply to these other topologies as well, and it would be important to study the lowest-action contributions to the no-boundary state. It would be interesting to understand if the analogues of Polchinski's phase appearing in the gravity partition function on a product of spheres~\cite{Shi:2025amq,Ivo:2025yek} are not only connected to instabilities in the spectrum of excitations of the static patch~\cite{Ivo:2025yek} (upon continuing one of the spheres to real time), but also to non-normalizable directions of the late-time wavefunction.
\\ \\
\textbf{Reviving Colemania.} One of Polchinski's motivations in the computation of his phase~\cite{Polchinski:1988ua} of the sphere partition function was to refine Coleman's proposal for the suppression of a (positive) cosmological constant~\cite{Coleman:1988tj}, which he termed `Colemania.'  In brief, Coleman's argument (streamlined in~\cite{Klebanov:1988eh}) is that contributions from small wormholes serve to induce a distribution over coupling constants in gravity, including the gravitational interaction and the cosmological constant. Then the norm of the no-boundary state, assumed by Coleman to be equal to the sphere partition function, at tree level pressures the induced distribution over the cosmological constant toward zero. 

Polchinski's phase~\cite{Polchinski:1988ua} appears to put a wrinkle in Coleman's argument; in $d = 4$ the phase is negative, and so one might jump to the conclusion that the cosmological constant should be driven to infinity instead of to zero.  Our calculations instead suggest a different conclusion: because Coleman's argument hinges on the norm of the no-boundary state rather than the sphere partition function, the finiteness and positivity of the norm (possibly due to other topologies at future infinity) would revive Colemania.  Of course other implications of and objections to Coleman's proposal of course need to be taken into account (see e.g.~comments in~\cite{Klebanov:1988eh}).

Here we also give an alternative argument, assuming for the moment that the leading contribution to the norm is positive. Suppose that $\Psi[\gamma]$ is the leading contribution to the no-boundary state.  In a third-quantized notation, we can write $\Psi[\gamma] = \langle \! \langle \gamma | a^\dagger |\Omega \rangle \! \rangle$ where $a^\dagger$ generates a single universe on top of the third-quantized vacuum $|\Omega\rangle$.  Since the universes are bosonic, the $n$-universe contribution is $\frac{1}{\sqrt{n!}} (a^\dagger)^n |\Omega\rangle$ and summing over all $n$ (including the $n = 0$ `no-universe' contribution) we find the leading contributions to the no-boundary state.  The norm goes schematically as $\sim \exp(\exp(S))$ where $S \sim 1/(G\Lambda)$ in four dimensions.  If we imagine that $\sim \exp(\exp(1/(G\Lambda)))$ induces a distribution over $\Lambda$'s by some mechanism related to Coleman's or otherwise, then the most likely value of the cosmological constant is the smallest nonzero one appearing in the distribution. (Of course, we really want $\Lambda$ to be finite but small; this would require some other mechanism.)  Since the norm is positive, loop corrections will not spoil the argument.  This argument also implies that the average `number of universes' is $\sim e^{S}$.
\\ \\
 \textbf{The $\ell = 0$ problem.} As recently reviewed in~\cite{Maldacena:2024uhs}, the no-boundary state coupled to a slow-roll inflaton has an action which pressures the inflaton to be near the bottom of its potential and thus enacts a small number of e-folds of inflation, predicting a vastly smaller spatial curvature for the universe than is observed.  The spatial curvature is tied to the $\ell = 0$ fluctuations of the no-boundary state coupled to the inflaton.  Our work shows that since the leading contribution to the no-boundary state (with a slow-roll inflaton) is null, the $\ell = 0$ problem should be re-examined for other contributions including other topologies.

In recent work,~\cite{Ivo:2024ill} studied the `no-boundary density matrix' of the universe, in which there is a manifestation of the same $\ell = 0$ problem coming from the leading saddle which is due to the Hartle-Hawking geometry.  They identify a subleading saddle related to the Coleman-de Luccia instanton which does not have the $\ell = 0$ problem in itself, but they state that this does not fully solve the $\ell = 0$ problem since the Hartle-Hawking geometry is the leading contribution.  If the no-boundary state is null, then the subleading saddle of~\cite{Ivo:2024ill} may be a candidate for a de facto leading saddle of the no-boundary density matrix which does not have the $\ell = 0$ problem.
\\ \\
\textbf{Conformal gravity at future infinity.} We have shown that unnormalized correlators of the no-boundary state can be computed by a theory of Euclidean conformal gravity at future infinity, consistent with the picture of~\cite{Strominger:2001pn, Maldacena:2002vr, Maldacena:2011mk}.  For instance, for pure 3d de Sitter gravity the theory at future infinity behaves like a bosonic world sheet string theory, aligning with recent work~\cite{Collier:2025lux}.  This point of view further supports the need to sum over topologies.
\\ \\
\textbf{A better definition of horizon entropy?}   We have shown that the norm of the no-boundary state is not the sphere partition function.  Although the complex time contour for the leading contribution to the norm of the no-boundary state is homologous to the complex time contour of the sphere, the reality conditions on fields are different, as evinced by the appearance of a residual symmetry of $SO(d,1)$ in the norm versus $SO(d+1)$ for the sphere.  Even though Gibbons and Hawking proposed that the sphere encodes the number of states in the static patch~\cite{Gibbons:1976ue}, the physical meaning of the sphere remains unclear due to Polchinski's phase~\cite{Polchinski:1988ua}.  Said another way, we do not have a clear grasp of the connection between the Euclidean computation of the sphere and de Sitter physics in real time.

We propose that the norm of the no-boundary state provides a better definition of the de Sitter horizon entropy, supplanting the usual Euclidean sphere. By coupling to a worldline with a clock as in~\cite{Witten:2023xze}, the norm gives a finite answer going as 
$\frac{\widetilde{Z}' S_0^{-(\mathcal{D}_{d-2} +1)/2}}{\text{vol}(SO(d-1))}\,e^{S_0 - \beta_{\text{dS}} m}$ which is likewise positive, reproduces the tree-level de Sitter horizon entropy in the presence of an observer, and has one-loop corrections similar to the sphere result.  As such, it is sensible to suggest that the norm of the no-boundary state might be `the' de Sitter horizon entropy, at least in the presence of an observer. 
\\ \\
\textbf{A speculation about other pairings.} We have argued above that the norm of the no-boundary state may be a better definition of the de Sitter horizon entropy. Here we speculate on how the ordinary sphere partition function may arise from a different kind of pairing. Suppose $\textsf{K}$ is a complex conjugation operator that acts on the asymptotic Hilbert space of states in de Sitter, so that e.g.
\begin{align}
\langle \! \langle  \gamma | \textsf{K}\, \text{HH} \rangle \! \rangle = \Psi_{\text{HH}}^*[\gamma]\,.
\end{align}
Then consider the three pairings, namely $\langle \! \langle \text{HH}  | \text{HH} \rangle \! \rangle$, $\langle \!\langle \textsf{K} \,\text{HH} |  \text{HH}  \rangle \! \rangle$, and $\langle \! \langle \text{HH}| \textsf{K} \,\text{HH} \rangle \! \rangle$.  The first is a bra-ket, the second is a ket-ket, and the third is a bra-bra.  In this paper we have shown that $\langle \! \langle \text{HH}  | \text{HH} \rangle \! \rangle 
\propto e^{S_0}$, and one may wonder if there is a sensible computation that establishes
\begin{align}
\langle \!\langle \textsf{K} \,\text{HH} |  \text{HH}  \rangle \! \rangle 
\stackrel{\text{?}}{\propto} i^{\,d+2}\;e^{S_0}\,,\quad
\langle \! \langle \text{HH}| \textsf{K} \,\text{HH} \rangle \! \rangle 
\stackrel{\text{?}}{\propto}  (-i)^{\,d+2}\;e^{S_0}\,.
\end{align}
 At present, we only fully understand how to compute the bra-ket; as discussed in this manuscript, the one-loop result is parametrically the absolute value of the sphere partition function divided by $\text{vol}(SO(d,1))$. We speculate here that the ket-ket may carry factors of $i$ and optimistically would align with Polchinski's $i$'s; then the bra-bra would follow similarly but with $i \to -i$.  It is tempting to suggest that the Euclidean sphere is the ket-ket.  These statements are resonant with recent discussions of inner products in field theory and gauge theory in~\cite{Witten:2025ayw}, in particular the suggestion that the Euclidean gravity path integral computes a pairing of kets with kets. In order to match the sphere, this pairing would involve a different reality condition on the late-time metric data leading to a division by the isometry group $SO(d+1)$ rather than $SO(d,1)$.
\\ \\
\textbf{Canonical formalism.} It would be nice to understand our inner product and the ensuing norm of the no-boundary state from the perspective of the canonical formalism, building off of e.g.~\cite{Higuchi:1991tk, Higuchi:1991tm}; see also~\cite{Marolf:2008hg, Marolf:2012kh} and in particular~\cite{Chakraborty:2023yed, Chakraborty:2023los} for more recent work.  For a recent discussion of the canonical formalism and inner products in the context of de Sitter JT gravity, see~\cite{Held:2024rmg}.
\\ \\
\indent Finally, we note that our approach to computing the norm of the no-boundary state introduces methods for evaluating loop-level inner products of wavefunctionals at asymptotia in asymptotically de Sitter spacetimes.  A systematic investigation of these inner products in 3d, as well as 4d and higher dimensions, will appear in forthcoming work~\cite{CJWIP1, CHJWIP2}.  Our perspective is that a careful treatment of the gravity path integral will reveal new features and subtleties of quantum gravity in de Sitter and so clarify our understanding of quantum cosmology.

\subsection*{Acknowledgements}

We thank Dionysios Anninos, Victor Gorbenko, Tom Hartman, William Harvey, Austin Joyce, Juan Maldacena, Suvrat Raju, Joaquin Turiaci, and Zhenbin Yang for valuable discussions. JC is supported by the Simons Collaboration on Celestial Holography, and by an Alfred P.~Sloan Foundation Fellowship. KJ is supported in part by an NSERC Discovery grant. KJ gratefully acknowledges support from the Simons Center for Geometry and Physics and the organizers of the program ``Black hole physics from strongly coupled thermal dynamics,'' where some of this research was performed.

\appendix

\section{Compare and contrast: the tachyonic scalar}
\label{App}

The conformal mode of gravity with positive cosmological constant behaves in many ways like a tachyonic scalar.  However, outside of a minisuperspace truncation, this association is merely a mnemonic.  For example, the no-boundary wavefunction $\langle \! \langle \gamma | \text{HH} \rangle \! \rangle = \Psi_{\rm HH}[\gamma]$ only depends on the boundary metric $\gamma$ and not the late-time value of the conformal mode  As such, there are no wrong-sign directions in the no-boundary state coming from the conformal mode.

In this Appendix, we perform an analysis of the Bunch-Davies wavefunction of a tachyonic scalar, so as to better calibrate our intuition about our gravity results.

Let us begin with the sphere partition function for a tachyon on a unit Euclidean sphere. The Euclidean action is
\beq
	S_E = \frac{1}{2}\int d^dx \sqrt{g} ((\partial\phi)^2 + M^2 \phi^2)\,,
\eeq
and we parameterize the mass as $M^2 = \left( \frac{d-1}{2}\right)^2 + \nu^2$ with $\nu = i \left( \tilde{\nu} + \frac{d-1}{2}\right)$. The tachyonic scalar corresponds to $\tilde{\nu}$ real and positive. In that case the sphere partition function does not converge with a real integration contour for $\phi$. We can assign it a finite value in the following way. Decomposing $\phi$ into spherical harmonics $Y^{(d)}_{j \textbf{n}}$ as $\phi = \sum_{j, \textbf{n}} \phi_{j \textbf{n}} Y^{(d)}_{j \textbf{n}}$ satisfying $-\nabla^2 Y^{(d)}_{j \textbf{n}} = j(j+d-1) Y^{(d)}_{j \textbf{n}}$, the modes of low $j< \tilde{\nu}$ appear in wrong-sign Gaussian integrals, while those with $j>\tilde{\nu}$ appear in right-sign Gaussian integrals. For generic $M^2$ so that there are no zero modes we proceed by rotating the contour for the low-lying modes with $j<\tilde{\nu}$. We can do so in two ways by rotating $\phi_{j<\tilde{\nu}} \to \pm i \phi_{j<\tilde{\nu}}$,\footnote{In a non-Gaussian theory whose action is bounded below the analogous procedure is to deform the contour of field integration to pass through the saddle point while following the steepest descent contour. } leading to a finite result upon renormalization,
\beq
	Z_{\rm sphere} = (\pm i)^{d_c} \text{det}^{-1/2}|-\nabla^2+M^2|\,,   
\eeq
where $d_c$ counts the number of modes with $j<\tilde{\nu}$. For example, when $-d<M^2 <0$ there is one unstable mode and $d_c =1$, while in the range $-2(d+1)<M^2<-d$ the $j=0$ and $j=1$ modes are unstable leading to $d_c = d+2$. The determinant is real and positive and so $Z_{\rm sphere}$ has the phase $(\pm i)^{d_c}$. This is analogous to Polchinski's phase insofar as the rotated conformal mode behaves to one-loop order on the sphere like a free scalar with $M^2 = -2(d-1)$.

Now imagine cutting the sphere along the equator. The Euclidean path integral on the hemisphere prepares the hemisphere Bunch-Davies state which we denote by $|\widetilde{\text{BD}}\rangle$.  If we evolve this state in Lorentzian de Sitter to future asymptotia, we obtain the usual Bunch-Davies state denoted by $|\text{BD}\rangle$.  Then the two are related by $|\text{BD}\rangle = \hat{U} |\widetilde{\text{BD}}\rangle$, where $U$ enacts time evolution from $t = 0$ to asymptotia.

We will show how to recover the phase $(\pm i)^{d_c}$ of the sphere partition function from both the wavefunction on the boundary of a Euclidean hemisphere $|\widetilde{\text{BD}}\rangle$ and from real-time evolution of the Bunch-Davies state in a rigid de Sitter space $|\text{BD}\rangle$. 

\subsection{Life at the equator}

It will be useful to first discuss the wavefunction for a stable matter field with $M^2>0$, i.e.~real $\nu$ or imaginary $\nu$ in the interval $0\leq -i\nu \leq \frac{d-1}{2}$, and then circle back to a tachyonic matter field later. For a stable field the sphere partition function equals the norm of the hemisphere Bunch-Davies state. The hemisphere Bunch-Davies wavefunction $\Psi_{\widetilde{\text{BD}}}[\tilde{\varphi}]$ of the latter is calculated by the path integral on a Euclidean hemisphere 
\beq
	ds^2 = d\tau^2 + \cos^2(\tau)d\Omega_{d-1}^2\,,
\eeq
with $\frac{\pi}{2}\geq \tau \geq 0$ with Dirichlet boundary conditions fixing $\phi(\tau=0,\Omega_{d-1}) = \tilde{\varphi}(\Omega_{d-1})$ on the equator. In an equation,
\beq
	\Psi_{\widetilde{\text{BD}}}[\tilde{\varphi}] = \int^{\tilde{\varphi}} [d\phi] e^{-S_E}\,.
\eeq
where the integral is taken over configurations that are regular at the pole $\tau = 0$ and obey the boundary condition on the equator. The action is $S_E$, rather than $S_E$ plus a boundary term, so as to be consistent with the Dirichlet condition. For real $\tilde{\varphi}$ the hemisphere Bunch-Davies wavefunction is real, and so in terms of $\Psi_{\rm BD}$ the sphere partition function is
\beq
	Z_{\rm sphere} =\langle \widetilde{\text{BD}} | \widetilde{\text{BD}}\rangle =  \int [d\tilde{\varphi}] \Psi_{\widetilde{\text{BD}}}[\tilde{\varphi}]^2\,.
\eeq

To calculate the wavefunction we separate $\phi$ into a classical trajectory obeying the boundary conditions and a fluctuation,
\beq
	\phi = \phi_0 + \delta\phi\,,
\eeq
where $\delta\phi$ vanishes at the equator and is regular at the pole $\tau = 0$. We can decompose $\delta\phi$ into suitable spherical harmonics, which we take to have definite angular momentum $j$ on the $d$-sphere and angular momentum $\ell$ on the $d-1$-sphere with $j\geq \ell$ and $j + \ell$ odd. Decomposing the boundary condition $\tilde{\varphi}$ into $d-1$-dimensional spherical harmonics
\beq
	\tilde{\varphi} (\Omega_{d-1}) = \sum_{\ell, \textbf{m}} \tilde{\varphi}_{\ell \textbf{m}}Y^{(d-1)}_{\ell \textbf{m}}(\Omega_{d-1})\,,
\eeq
we take $\phi_0$ to be the regular solution to the Klein-Gordon equation obeying the boundary condition, 
\begin{align}
\begin{split}
\label{E:classicalBD1}
	\phi_0(\tau,\Omega_{d-1}) &=\sum_{\ell, \textbf{m}} \tilde{\varphi}_{\ell \textbf{m}}Y^{(d-1)}_{\ell \textbf{m}}(\Omega_{d-1}) \tilde{\psi}^{(d)}_{\ell}(\tau)\,,
	\\
	\tilde{\psi}^{(d)}_{\ell}(\tau) &= \widetilde{C}^{(d)}_{\ell}  \cos^{\ell}(\tau) _2F_1 \left( \frac{d-1+2(\ell-i\nu)}{4},\frac{d-1+2(\ell+i\nu)}{4};\frac{d}{2}+\ell;\cos^2(\tau)\right)\,, 
	\\
	 \widetilde{C}^{(d)}_{\ell}& = \frac{\Gamma\left( \frac{d+1+2(\ell-i\nu)}{2}\right)\Gamma\left( \frac{d+1+2(\ell+i\nu)}{2}\right)}{\sqrt{\pi}\Gamma\left( \frac{d}{2}+\ell\right)}\,.
\end{split}
\end{align}
For positive $\tau$ near the equator the mode functions $\psi^{(d)}_{\ell}$ behave as
\beq
	\tilde{\psi}^{(d)}_{\ell}(\tau)= 1 - \frac{2 \Gamma\left( \frac{d+1+2(\ell-i\nu)}{4}\right)\Gamma\left( \frac{d+1+2(\ell+i\nu)}{4}\right)}{\Gamma\left( \frac{d-1+2(\ell-i\nu)}{4}\right)\Gamma\left( \frac{d-1+2(\ell+i\nu)}{4}\right)}\tau + O(\tau^2)\,.
\eeq
The action of the classical configuration $\phi_0$ is a boundary term
\beq
	S_E[\phi_0] =-\frac{1}{2}\int_{\tau=0} d\Omega_{d-1} \,\phi_0 \partial_{\tau}\phi_0 = \sum_{\ell, \textbf{m}} \frac{\Gamma\left( \frac{d+1+2(\ell-i\nu)}{4}\right)\Gamma\left( \frac{d+1+2(\ell+i\nu)}{4}\right)}{\Gamma\left( \frac{d-1+2(\ell-i\nu)}{4}\right)\Gamma\left( \frac{d-1+2(\ell+i\nu)}{4}\right)} |\tilde{\varphi}_{\ell \textbf{m}}|^2 = \sum_{\ell, \textbf{m}} \widetilde{\Upsilon}_{\ell}^{(d)}|\tilde{\varphi}_{\ell \textbf{m}}|^2\,,
\eeq
giving a Gaussian wavefunction
\beq
	\Psi_{\widetilde{\text{BD}}}[\tilde{\varphi}] = \widetilde{\Psi}_{1\text{-loop}}\exp\!\left( -\sum_{\ell, \textbf{m}} \widetilde{\Upsilon}_{\ell}^{(d)}|\tilde{\varphi}_{\ell \textbf{m}}|^2\right)\,, \qquad \widetilde{\Psi}_{1\text{-loop}} = \widetilde{\text{det}}^{-1/2}(-\nabla^2+M^2)\,,
\eeq
where the tilde over the determinant indicates that it is taken over those modes obeying the boundary condition, i.e.~$j\geq \ell\geq 0$ with $j+\ell$ odd. Because $\nu$ is imaginary this wavefunction is a real, right-sign Gaussian in all directions $\tilde{\varphi}_{\ell}$,\footnote{The reality of the wavefunction is tantamount to saying that we can interpret the Euclidean path integral as preparing a bra equally well as a ket.} and it is an instructive exercise to demonstrate that this state is not normalized to unity, but rather the sphere partition function.

Now consider the tachyonic scalar which corresponds to $\tilde{\nu} = -i \nu - \frac{d-1}{2}$ real and positive. As for the stable scalar we can decompose $\phi$ into a classical trajectory $\phi_0$ plus fluctuations $\delta\phi$, the latter of which are identical to before. Those fluctuations with $j < \tilde{\nu}$ are wrong-sign and must be rotated to assign a finite value to the wavefunction. We let $\tilde{d}_c$ denote the number of such wrong-sign fluctuations. For $\tilde{\nu}<1$ the constraint $j+\ell $ odd together with $j\geq \ell$ forbid such wrong-sign fluctuations, i.e.~$\tilde{d}_c=0$ and the hemisphere Bunch-Davies wavefunction is unambiguous in that case, while in the range $1<\tilde{\nu}<2$, i.e.~$-2(d+1)<M^2 <-d$, only the $j=1$ mode with $\ell=0$ is wrong-sign and must be rotated giving $\tilde{d}_c=1$. In general the hemisphere Bunch-Davies wavefunction is no longer real and there are two versions of the Euclidean path integral, one that prepares the wavefunction say with the rotations by a $+i$, and the other its conjugate.

Meanwhile, the classical trajectory $\phi_0$ is identical to that of a stable scalar. In terms of $\tilde{\nu}$ the coefficients $\Upsilon^{(d)}_{\ell}$ that determine the variances of the hemisphere Bunch-Davies wavefunction become
\beq
	\widetilde{\Upsilon}^{(d)}_{\ell} = \frac{\Gamma\left( \frac{d+\ell+\tilde{\nu}}{2}\right) \Gamma\left( \frac{\ell+1- \tilde{\nu}}{2}\right)}{\Gamma\left( \frac{d-1+\ell+\tilde{\nu}}{2}\right)\Gamma\left( \frac{\ell-\tilde{\nu}}{2}\right)}\,.
\eeq
We then have
\beq
	\Psi_{\widetilde{\text{BD}}}[\tilde{\varphi}] = \widetilde{\Psi}_{1\text{-loop}} \exp\!\left( - \sum_{\ell, \textbf{m}}\Upsilon_{\ell}^{(d)}|\tilde{\varphi}_{\ell \textbf{m}}|^2\right)\,, \qquad \widetilde{\Psi}_{1\text{-loop}} = (\pm i)^{\tilde{d}_c}\widetilde{\text{det}}^{-1/2}|-\nabla^2+M^2|\,.
\eeq

For the tachyonic scalar the $\widetilde{\Upsilon}^{(d)}_{\ell}$'s are real but at least one is negative. For example in the range $0<\tilde{\nu}<1$, $\widetilde{\Upsilon}^{(d)}_{\ell=0}$ is negative while the $\widetilde{\Upsilon}^{(d)}_{\ell>0}$ are positive. As another example, in the range $1<\tilde{\nu}<2$, the $\widetilde{\Upsilon}^{(d)}_{\ell=0}$ and $\widetilde{\Upsilon}^{(d)}_{\ell>1}$ are positive while $\widetilde{\Upsilon}^{(d)}_{\ell=1}$ is negative. In the former case this means that the hemisphere Bunch-Davies wavefunction is a wrong-sign Gaussian distribution in the $\ell=0$ fluctuations of $\tilde{\varphi}$ but right-sign in the $\ell>0$ directions. In the latter, it is a right-sign Gaussian in the $\ell=0$ and $\ell>1$ fluctuations, but a wrong-sign Gaussian in the $\ell=1$ directions. Let $D_c$ denote the number of such wrong-sign directions.

The sphere partition function of the tachyonic scalar is related to the ``norm'' of its hemisphere Bunch-Davies wavefunction in the following way. (Here and throughout we implicitly have the finite and in general complex values which we assign to the sphere partition function and the wavefunction.) Rotating the wrong-sign fluctuations $\tilde{\varphi}_{\ell \textbf{m}}$ of the wavefunction, we still have
\beq
	Z_{\rm sphere} = \int [d\tilde{\varphi}] \Psi_{\widetilde{\text{BD}}}[\tilde{\varphi}]^2\,,
\eeq
i.e.~we integrate the square of the wavefunction rather than its complex square. The LHS has a phase we discussed previously, $(\pm i)^{d_c}$ with $d_c$ the number of wrong-sign modes on the sphere with $j<\tilde{\nu}$. Provided that we always rotate the wrong-sign modes appearing in the prefactor of $\Psi_{\widetilde{\text{BD}}}$ and its argument in the same way, i.e.~with a $i$ or a $-i$, the RHS has the same phase, $(\pm i)^{2\tilde{d}_c + D_c}$ with $d_c = 2\tilde{d}_c + D_c$. We summarize the counting problem in Table~\ref{T:tachyonicJoe}.

\begin{table}[h]
\begin{center}
\begin{tabular}{c|c|c|c}
	 &$ 0<\tilde{\nu}<1$ & $ 1<\tilde{\nu}<2$ & $ 2<\tilde{\nu}<3$ 
	\\
	\hline
	$d_c$ & 1 & $d+2$ & $\frac{1}{2}(d+4)(d+1)$
	\\
	\hline
	$\tilde{d}_c$ & 0 & 1 & $d+1$ 
	\\
	\hline
	$D_c$ & 1 & $d$ & $\frac{d(d+1)}{2}$
\end{tabular}
\end{center}
\caption{\label{T:tachyonicJoe} Counting the number $d_c$ of wrong-sign fluctuations of the sphere partition function, the number $\tilde{d}_c$ of the hemisphere Bunch-Davies wavefunction, and the number $D_c$ of its argument $\tilde{\varphi}_{\ell}$, for various ranges of $\tilde{\nu}$. In each case $d_c=2\tilde{d}_c+D_c$.}
\end{table}

Note that if we integrated $|\Psi_{\rm BD}|^2$ instead of the ordinary square, then the ``norm'' still has a phase $(\pm i)^{D_c}$ which is almost that of the sphere, since $d_c \equiv D_c \,\,(\text{mod }2)$.

The broad lesson is that in this simple example the appearance of wrong-sign modes on the sphere corresponds to a sickness of the Lorentzian theory, namely the non-normalizability of the hemisphere Bunch-Davies state. In a non-Gaussian theory where the Euclidean action is bounded below the corresponding computations would indicate that $\tilde{\varphi} = 0$ is an unstable saddle point of the probability distribution for $\tilde{\varphi}$ on the equator. 

\subsection{Life at future infinity}

Now let us take the hemisphere Bunch-Davies state on the equator $|\widetilde{\text{BD}}\rangle$ and evolve it in real time where the sphere continues to a part of de Sitter space, in particular to late real time. For the stable scalar, by unitarity of evolution we can evaluate the norm at any real time we wish and the norm will remain the sphere partition function. For the tachyonic scalar we will see that a similar statement holds true up to phases. 

The late-time wavefunction $|\text{BD}\rangle$ is computed by a path integral for the scalar on the spacetime~\eqref{E:globaldS} with the complex time contour described there. We integrate over scalar configurations obeying two boundary conditions, regularity at the north pole of the Euclidean hemisphere, and we suitably fix the field at late time. For simplicity let us suppose that the scalar is light, with $M^2$ in the range $0<M^2<\frac{(d-1)^2}{4}$, i.e.~$\nu$ is imaginary in the interval $ 0 < -i \nu < \frac{d-1}{2}$. Then classically $\phi$ has the late-time behavior
\beq
	\phi(t,\Omega_{d-1}) \sim e^{(\Delta - d+1)t} \varphi(\Omega_{d-1}) + e^{-\Delta t} \zeta(\Omega_{d-1})\,, \qquad \Delta = \frac{d-1}{2} - i \nu\,,
\eeq
where this mass range implies $\frac{d-1}{2}< \Delta < d-1$. Fixing $\phi$ as $t\to\infty$ is tantamount to fixing the coefficient $\varphi$ of the leading decay, allowing terms at most as big as $e^{-\Delta t}$ to fluctuate. The wavefunction is
\beq
	\Psi_{\rm BD}[\varphi] = \langle \varphi|\text{BD}\rangle = \int^{\varphi} [d\phi] \, e^{iS}\,, \qquad S =-\frac{1}{2} \int d^dx \sqrt{-g} \left( (\partial\phi)^2 + M^2 \phi^2\right) + (\text{bdy})\,,
\eeq
and we tune the boundary terms so that the action and so also the wavefunction has a well-defined late time limit. 

To calculate the wavefunction one again separates $\phi$ into a classical trajectory $\phi_0$ that obeys the late-time boundary condition, and (in this case, a continuum of) normalizable fluctuations $\delta\phi$. The wavefunction becomes
\beq
	\Psi_{\rm BD}[\varphi] = \Psi_{1\text{-loop}} e^{iS[\phi_0]}\,, \qquad \Psi_{1\text{-loop}} = \text{det}^{-1/2}(-\nabla^2 + M^2)\,,
\eeq
where the one-loop prefactor comes the integral over fluctuations. To calculate the classical trajectory one decomposes $\varphi$ into angular harmonics on the late-time sphere,
\beq
	\varphi(\Omega_{d-1}) = \sum_{\ell, \textbf{m}} \varphi_{\ell \textbf{m}} Y^{(d-1)}_{\ell \textbf{m}}(\Omega_{d-1})\,,
\eeq
and then 
\begin{align}
\begin{split}
	\phi_0(t,\Omega_{d-1}) &= \sum_{\ell, \textbf{m}} \varphi_{\ell \textbf{m}} \psi_{\ell}^{(d)}(t) Y^{(d-1)}_{\ell \textbf{m}}(\Omega_{d-1})\,,
	\\
	\psi^{(d)}_{\ell}(t)& = C_{\ell} \cosh^{\ell}(t) _2F_1\left( \frac{d-1+2(\ell-i \nu)}{4},\frac{d-1+2(\ell+i \nu)}{4};\frac{d}{2}+\ell;\cosh^2(t)\right) 
	\\
	& = \frac{C_{\ell}}{\widetilde{C}_{\ell}} \tilde{\psi}^{(d)}_{\ell}(\tau = -it)\,,
	\\
	C_{\ell} & = \frac{\Gamma\left( \frac{d+1+2(\ell-i\nu)}{4}\right)\Gamma\left( \frac{d-1+2(\ell-i \nu)}{4}\right)}{\Gamma\left( \frac{d}{2}+\ell\right)\Gamma\left( -i\nu\right) } \exp\left( -\frac{i \pi}{2}\left( \frac{d-1+2(\ell+i \nu)}{2}\right)\right)\,,
\end{split}
\end{align}
where the mode function $\psi_{\ell}^{(d)}(t)$ is proportional to a certain continuation of the mode $\tilde{\psi}^{(d)}_{\ell}(\tau)$ in~\eqref{E:classicalBD1} on the hemisphere to real time. In particular we must approach $t\to\infty$ slightly above the real axis. The late-time limit of $\psi$ is 
\beq
	\psi^{(d)}_{\ell}(t) \sim e^{(\Delta-d+1)t} + e^{-\Delta t} \frac{\Gamma\left( \frac{d-1+2(\ell-i \nu)}{2}\right)\Gamma(i\nu)}{\Gamma\left( \frac{d-1+2(\ell+i \nu)}{2}\right)\Gamma(-i\nu)}e^{\pi \nu}\,.
\eeq

Without including boundary terms, the action of this trajectory integrated up to a time $t=T$ diverges as $T\to \infty$. This divergence is very similar to the large-volume divergences of the on-shell action of quantum field theory in anti-de Sitter space, for which the remedy is holographic renormalization. In the late time limit of de Sitter we can play a similar game, integrating the action up to a finite but large time $T$, adding local counterterms on the $t=T$ slice built from the boundary data with real coefficients which can be thought of as stripping off large phases from $e^{iS}$, and then taking the $T\to\infty$ limit. This procedure results in a ``renormalized'' action $S$, leaves the probability distribution $|\Psi_{\rm BD}|^2$ invariant, and in this setting yields 
\beq
	S[\phi_0]  =   -2^{-d}\Delta\sum_{\ell, \textbf{m}} \frac{\Gamma\left( \frac{d-1+2(\ell-i \nu)}{2}\right)\Gamma(i\nu)}{\Gamma\left( \frac{d-1+2(\ell+i \nu)}{2}\right)\Gamma(-i\nu)}e^{\pi \nu}|\varphi_{\ell \textbf{m}}|^2 = i \sum_{\ell, \textbf{m}} \Upsilon_{\ell}^{(d)} |\varphi_{\ell \textbf{m}}|^2\,.
\eeq
This leads to a probability distribution on $\varphi$,
\beq
	|\Psi_{\rm BD}[\varphi]|^2 = |\Psi_{1\text{-loop}}|^2 \exp\!\left(-2\sum_{\ell, \textbf{m}} \text{Re}(\Upsilon^{(d)}_{\ell})|\varphi_{\ell \textbf{m}}|^2\right)\,.
\eeq
This is a joint Gaussian distribution in the $\varphi_{\ell \textbf{m}}$'s, and for a stable scalar all of the $\text{Re}(\Upsilon)$'s are positive so that the distribution is normalizable. The ultralocal inner product of states of definite $\phi$ becomes an ultralocal inner product for $\varphi$, namely $\langle \varphi|\varphi'\rangle = \delta[\varphi-\varphi']$, and so by unitarity of evolution
\beq
	Z_{\rm sphere} = \int [d\varphi] |\Psi_{\rm BD}[\varphi]|^2\,.
\eeq

As with our discussion of the Bunch-Davies wavefunction on the equator, almost all of these statements continue to go through for a tachyonic scalar for which $-i \nu > \frac{d-1}{2}$. In particular the classical trajectory is the same as that above and so the wavefunction gives rise to a joint Gaussian distribution for the $\varphi_{\ell \textbf{m}}$'s,
\beq
	|\Psi_{\rm BD}[\varphi]|^2 \propto \exp\!\left( - 2 \sum_{\ell, \textbf{m}}\text{Re}(\Upsilon^{(d)}_{\ell}) |\varphi_{\ell \textbf{m}}|^2\right),
\eeq
only now some of the directions are wrong-sign Gaussians depending on the mass-squared in exactly the same way as for the wavefunction on the equator. For $-d<M^2<0$, i.e.~$0<\tilde{\nu}<1$, only the $\ell=0$ mode of $\varphi$ has wrong-sign distribution, while in the range $-2(d+1)<M^2<-d$, i.e.~$1<\tilde{\nu}<2$, the $\ell=1$ modes of $\varphi$ have a wrong-sign distribution. To compute the ``norm'' of the late-time state in that case one could rotate the wrong-sign modes by $90^{\circ}$ at the cost of the same phase $(\pm i)^{D_c}$ encountered on the equator.

\bibliography{refs}
\bibliographystyle{JHEP}

\end{document}